\documentclass[final,times,natbib,floatfix]{elsarticle}
\usepackage{bm}
\usepackage{amsmath}
\usepackage{epsfig}
\usepackage{rotating}
\usepackage{graphicx}
\usepackage{url}

\def\pr{\prime}

\def\be{\begin{equation}}
\def\lan{\left\langle}
\def\ran{\right\rangle}
\def\ee{\end{equation}}
\def\barr{\begin{array}}
\def\earr{\end{array}}

\def\l{\left}
\def\r{\right}
\def\f{\frac}
\def\dis{\displaystyle}
\def\ed{\end{document}}
\def\la{\lambda}

\def\can{{\cal N}}

\def\cc{{\cal C}}

\def\baa{{\mbox{\boldmath $\alpha$}}}

\begin{document}
\sloppy

\journal{Nuclear Physics A}

\begin{frontmatter}

\title{Multiple $SU(3)$ algebras in interacting boson model and shell model: Results for $(\beta ,\gamma$) bands and scissors $1^+$ band}

\author[label4]{V.K.B. Kota}
\ead{vkbkota@prl.res.in}
\author[label3]{R. Sahu}
\ead{rankasahu@gmail.com}

\address[label4]{Physical Research Laboratory, Ahmedabad 380 009, India}
\address[label3]{National Institute of Science and Technology, Palur Hills, Berhampur-761008, Odisha, India}


\date{\hfill \today}

\begin{abstract}

Shell model (SM) and interacting boson model (IBM) spaces admit multiple $SU^{(\baa)}(3)$ algebras generating the same rotational spectra but different $E2$ decay properties, depending on the phases $\baa$ in the quadrupole generator. In the ground ($g$) $K=0^+$ bands in nuclei this is demonstrated recently using systems with nucleons in a single oscillator shell [Kota, Sahu and Srivastava, Bulg. J. Phys. {\bf 46}, 313 (2019); Eur. Phys. J. Special Topics {\bf 229}, 2389 (2020)]. Going beyond these preliminary studies, 
results are presented here for $E2$ decay properties of $\beta$ and $\gamma$ bands members, as generated by multiple $SU(3)$ algebras, using $sdg$IBM and $sdgi$IBM examples. Also, presented are some results for the $\gamma$ band using a SM example with eight protons in $sdg$ space. In addition, results are presented for the $E2$ and $M1$ decay properties of the levels of the scissors  $1^+$  band in heavy nuclei using $sdg$IBM-2 and $sdgi$IBM-2. The scissors $1^+$ band properties are also studied using a SM example with six protons in $(pf)$ shell and twelve neutrons in $(sdg)$ shell. These results establish that: (i) with multiple $SU(3)$ algebras, it is possible to have rotational bands with very weak $E2$ strengths among the levels where normally one expects strong strengths; (ii) $E2$ decay of the levels of $\gamma$ band (also $\beta$ band) to the ground band levels are quite different for some of the $SU^{(\baa)}(3)$ algebras with strong dependence on $(\baa)$; (iii) it is possible to have the scissors $1^+$ band with the $E2$ and $M1$ decay of the lowlying levels of this band to the $g$ band levels are strong or weak depending on $(\baa)$. 

\end{abstract}


\end{frontmatter}



\section{Introduction}

Elloitt's introduction of $SU(3)$ in nuclear shell model is a milestone in nuclear physics in describing quadrupole collective rotational states in nuclei from first principles \cite{El58-1,El58-2}. Similarly, $SU(3)$ is also central to the development of the interacting boson model \cite{Iac}. Within
the shell model (SM) context, $SU(3)$ also appears in the pseudo-$SU(3)$ model, proxy-$SU(3)$ scheme, fermion dynamical symmetry model, $Sp(6,R)$ model that includes multi-shell excitations and so on \cite{sm1,sm2,sm3,sm4,sm5,sm6,sm7,sm8,Kota-book}. Similarly,
within the interacting boson model (IBM) context, $SU(3)$ appears not just in IBM-1 but also in proton-neutron IBM or IBM-2, IBM-3 or isospin invariant IBM, IBM-4 or spin-isospin invariant IBM, $spdf$IBM, $sdg$IBM, interacting boson-fermion models for odd-A and odd-odd nuclei,  clustering models and so on \cite{Iac,Piet,ib2,ib3,ib4,ib5,ib6,ib7,ib8,KS-book}. The literature on $SU(3)$ in nuclei continues to expand with the opening of new directions in the applications of $SU(3)$ \cite{Kota-book}. One such new direction that is recognized recently is that both the SM and IBM admit multiple $SU(3)$ algebras \cite{Ko-17,KSP-19}. For example, given a oscillator major shell quantum number $\eta$ with fermions or bosons, there will be $2^{[\f{\eta}{2}]}$ number of $SU(3)$ algebras; $[\f{\eta}{2}]$ is the integer part of $\eta/2$. These arise due to various phase choices ($\baa$) possible (see Section 2 for details) in the quadrupole generator of $SU(3)$. Although the existence of two $SU(3)$ algebras in $sd$IBM-1 and in shell model $sd$ space are known before \cite{Iac,JCP}, multiple $SU(3)$ algebras, in the non trivial situations that go beyond the $sd$ space in IBM or SM, started receiving significant attention only from 2017 \cite{Ko-17,KSP-19,EPJST}. 

Investigating the structures generated by multiple $SU^{(\baa)}(3)$ algebras in SM and IBM, quadrupole deformed shapes (prolate or oblate) and $E2$ decay properties for systems with $SU(3)$ irreducible representations (irreps) of the type $(\lambda ,0)$ with $\lambda$ even giving the lowest $K=0^+$ band in even-even nuclei in $sdg$ space are reported in \cite{KSP-19,Kota-book} and in $sdgi$ space in \cite{EPJST,Kota-book}. Let us mention that $(\baa)$ takes four values in sdg space giving four $SU(3)$ algebras and similarly, there are eight $SU(3)$ algebras in $(sdgi)$ space. Let us mention that the $SU(3)$ irreps are in general labelled by two positive numbers $\lambda$ and $\mu$ and denoted by $(\lambda , \mu)$ or just $(\lambda \mu)$ when there is no confusion. Methods for obtaining $(\lambda \mu)$ in IBM and SM are given in \cite{Ko-irreps,Kota-book} with a simple formula for the lowest irrep. It is well known that in IBM, the lowest $SU(3)$ irrep is $(\eta N,0)$ where $N$ is boson number; for example $\eta=2$ for $sd$IBM, 4 for $sdg$IBM and 6 for $sdgi$IBM. The ($\baa$) dependence of various quadrupole properties are analyzed using the analytical formulation due to Kuyucak and Morrison \cite{KM-1} that is valid for sufficiently large values of $N$ and more importantly it applies to any ($\baa$). Results are presented in \cite{KSP-19,EPJST,Kota-book} for systems of 10 and 15 bosons in $sdg$IBM and $sdgi$IBM. Unlike in IBM, in SM the lowest $SU(3)$ irrep will be $(\lambda ,0)$ type with $\lambda$ even, for nuclei with even number of identical valence nucleons with total spin $S=0$ or for even-even nuclei with a given value of isospin $T$ and Wigner's $SU(4)$ irrep, only for some special nucleon (fermion) numbers; see \cite{Ko-irreps}. Calculations in \cite{KSP-19,EPJST} are restricted to these numbers. Within the shell model, besides employing shell model codes, in several examples used also is the deformed shell model (DSM) based on Hartree-Fock single particle states; details of DSM are given in \cite{KS-book}. With $p$ for protons and $n$ for neutrons, in \cite{KSP-19} analyzed are $(sdg)^{6p:S=0}$, $(sdg)^{6p,2n:S=0,T=2}$ and $(sdg)^{6p,6n:S=0,T=0}$ SM systems giving the lowest $SU(3)$ irreps to be $(18,0)$, $(26,0)$ and $(36,0)$ respectively. Similarly, in \cite{EPJST} analyzed are $(sdgi)^{6p:S=0}$, $(sdgi)^{6p,6n:S=0,T=0}$ and $(sdgi)^{12p,6n:S=0,T=3}$ SM systems giving the lowest $SU(3)$ irreps to be $(30,0)$, $(60,0)$ and $(78,0)$ respectively. These $SU(3)$ irreps follow easily from the formula given in \cite{Ko-irreps}. Also note that $6p$ stands for six protons, $6n$ stands for six neutrons and so on. A generic result obtained by these numerical investigations of multiple $SU(3)$ algebras in SM and IBM is that it is possible to have $SU(3)$ algebras generating rotational spectra with small quadrupole moments and weak $E2$ decay strengths. Another result is that shape will be prolate for most $(\baa)$ values and oblate for the remaining.  For further understanding and applications of multiple $SU^{(\baa)}(3)$ algebras, many more investigations are clearly needed and some of them are as follows.

\begin{enumerate}

\item Going beyond the ground $K=0^+$ band (hereafter called $g$ band) properties studied till now, clearly quadrupole transition strengths for the decay of levels of $\gamma$ and $\beta$ band to the levels of $g$ band in even-even nuclei in $sdg$IBM and $sdgi$IBM need to be analyzed. Similar studies in SM (using SM codes or DSM) are to be carried out for systems with the lowest $SU(3)$ irrep of the type $(\lambda ,2)$ with $\lambda$ even [this gives $K=0^+$ and $2^+$ ($\gamma$) bands] or $(\lambda ,4)$ with $\lambda$ even [this gives $K=0^+$, $2^+$ ($\gamma$) and $4^+$ bands]. Note that, in the SM description we have fermions and therefore in rare situations only, the lowest $SU(3)$ irrep will be $(\lambda ,0)$ type with $\lambda$ even giving a $K=0^+$ band \cite{Kota-book}.
 
\item Multiple $SU(3)$ algebras in IBM-2 and in the shell model context for nuclei with valence protons and neutrons in different shells (this is the situation with heavy nuclei) will allow one to analyze $M1$ and $E2$ decay properties of the levels of scissors $1^+$ band in heavy deformed nuclei. It is important to know their dependence on ($\baa$) of $SU^{(\baa)}(3)$.

\item Quantum phase transitions (QPT) and shape coexistence in nuclei using multiple $SU(3)$ algebras and also multiple pairing algebras studied in \cite{Ko-17} are expected to give new insights into QPT and here, multiple pairing plus quadrupole-quadrupole Hamiltonians may prove to be useful. See \cite{RMP} for investigations using multiple pairing [$SO(6)$] and $SU(3)$ algebras in $sd$IBM-1. Also, a closely related topic where multiple pairing and $SU(3)$ algebras play an important role is in generating order-chaos-order-$\dots$ transitions. \cite{RMP,chaos1,chaos2}.

\item Analysis of experimental data looking for signatures of multiple $SU^{(\baa)}(3)$ algebras need to be carried out. Also, it is important to identify some new experiments for testing the results due to multiple $SU(3)$ algebras. These will establish the role of $\baa$ in rotational nuclei across the periodic chart. 

\end{enumerate}

In the present work we will focus on items (1) and (2) and consider (3) and (4) in a separate publication. Now we will 
give a preview.

Section 2 gives a brief introduction to multiple $SU^{(\baa)}(3)$ algebras in SM and IBM and then presents results for $E2$ transition strengths from low-lying levels in $\gamma$ and $\beta$ bands to the $g$ band levels generated by multiple $SU^{(\baa)}(3)$ algebras in $sdg$IBM and $sdgi$IBM. In Section 3, presented are some SM results for $\gamma$ band using a system of eight protons in $(sdg)$ space. 
In Section 4, results are presented for the $E2$ and $M1$ decay properties of the levels of the scissors $1^+$ band in heavy nuclei using $sdg$IBM-2 and $sdgi$IBM-2. 
In Section 5 results for the scissors $1^+$ band in SM are presented for a system with valence protons in $(pf)$-shell and neutrons in $(sdg)$-shell. Finally, Section 6 gives conclusions. 

\section{Properties of $\gamma$ and $\beta$ bands in $sdg$IBM and $sdgi$IBM with multiple $SU(3)$ algebras}

In this Section we will restrict to interacting boson models $sdg$IBM and $sdgi$IBM where no distinction is made between proton and neutron bosons (called IBM-1's in literature).

\subsection{Multiple $SU(3)$ algebras}

Given an oscillator major shell with major shell quantum number $\eta$, the spectrum generating algebra (SGA) is $U(\can)$ with $\can = (\eta +1)(\eta +2)/2$. Also, for a given $\eta$, the 
orbital angular momentum $\ell$ of a single particle (it may be a boson as in IBM or a fermion as in SM) in the $\eta$ shell takes values $\ell = \eta$, $\eta-2$, $\ldots$, $0$ or $1$. Now, as Elliott has established, $U(\can) \supset SU(3) \supset SO(3)$ where $SO(3)$ generates orbital angular momentum. The eight generators of $SU(3)$ are the three angular momentum operators 
$$
L^1_q = \dis\sum_\ell \;\dis\sqrt{\ell(\ell+1)(2\ell+1)/3}
\;\l(b^\dagger_\ell \tilde{b}_\ell\r)^1_q 
$$
and the five quadrupole moment operators $Q^2_\mu(\baa)$ given by
\be
\barr{l}
Q^2_\mu(\baa) = \dis\sum_{\ell}\,t^{(\eta)}_{\ell , \ell} 
\l(b^\dagger_{\ell} \tilde{b}_{\ell}\r)^2_\mu + \dis\sum_{\ell_1\neq \ell_2}\,t^{(\eta)}_{\ell_1,\ell_2} 
\l(b^\dagger_{\ell_1} \tilde{b}_{\ell_2}\r)^2_\mu \\
= -(2\eta+3)\,\dis\sum_\ell
\dis\sqrt{\dis\frac{\ell(\ell+1)(2\ell+1)}{5(2\ell+3)(2\ell-1)}}
\l(b^\dagger_\ell \tilde{b}_\ell\r)^2_\mu \\
+ \dis\sum_{\ell <
\eta}\,\alpha_{\ell,\ell+2}\;\dis\sqrt{\dis\frac{6(\ell+1)(\ell+2)(\eta-\ell)
(\eta+\ell+3)}{5(2\ell+3)}}\l[\l(b^\dagger_\ell \tilde{b}_{\ell+2}\r)^2_\mu +
\l(b^\dagger_{\ell+2} \tilde{b}_{\ell}\r)^2_\mu \r]\;; \\
\alpha_{\ell_1,\ell_2}= \alpha_{\ell_2,\ell_1}\,,\;\;t^{(\eta)}_{\ell_1,\ell_2}= t^{(\eta)}_{\ell_2,\ell_1}\;,\\ 
\baa=(\alpha_{0,2}, \alpha_{2,4}, \ldots, \alpha_{\eta-2,\eta})\;\;
\mbox{for}\;\;\eta\;\;\mbox{even}\;,\\
\baa=(\alpha_{1,3}, \alpha_{3,5}, \ldots, \alpha_{\eta-2,\eta})\;\;
\mbox{for}\;\;\eta\;\;\mbox{odd}\;,\\
\baa=(\pm 1, \pm 1, \ldots)\;.
\earr \label{eq.new-1}
\ee
Here, $b^\dagger_{\ell, m}$ and $b_{\ell,m}$ are boson creation and annihilation operators, $m$ is $l_z$ quantum number and $\tilde{b}_{\ell,m} = (-1)^{\ell-m} b_{\ell , m}$. It is useful to mention that in the spectroscopic notation $b^\dagger_{\ell =0} = s^\dagger$, $b^\dagger_{2,m} = d^\dagger_m$, $b^\dagger_{4,m} = g^\dagger_m$, $b^\dagger_{6,m} = i^\dagger_m$ and so on. It is important to  
note that the formulas for $t^{(\eta)}_{\ell_1 , \ell_2}$ in Eq. (1) follow from the second and third lines in the equation. Also, $t^{\eta}_{\ell_1 , \ell_2}$
with $\ell_1 \neq \ell_2$ contain the phases $\alpha_{\ell,\ell+2}$. By evaluating the commutators $[L^1_q , L^1_{q^\prime}]$,
$[L^1_q , Q^2_\mu(\baa)]$ and $[Q^2_\mu(\baa) ,  Q^2_{\mu^\prime}(\baa)]$ it is easy to see that they form the $SU(3)$ algebra.    
Thus, for each choice of ($\baa$) in Eq. (\ref{eq.new-1}) there is a $SU^{(\baa)}(3)$ algebra. Clearly,
given a $\eta$, the number of $SU(3)$ algebras is $2^{\l[\f{\eta}{2}\r]}$ where
$\l[\f{\eta}{2}\r]$ is the integer part of $\eta/2$. Then, there will be two
$SU(3)$ algebras for $\eta=2$ shell ($sd$ space), four in $\eta=4$ shell ($sdg$
space), eight in $\eta=6$ shell ($sdgi$ space) and so on. This applies to both SM and IBM. In SM it is standard to use $\alpha_{\ell , \ell+2}=-1$ for all $\ell$ \cite{SM-MM1,SM-MM2} while in IBM it is standard to use $\alpha_{\ell , \ell+2}=+1$ \cite{Iac}. Our interest is in the study of the consequences of using all allowed phase choices 
for $\alpha_{\ell , \ell+2}$. 

With $N$ bosons in $sdg$IBM for example, the $g$ ($K^\pi = 0^+$) band is generated by the $SU(3)$ irrep $(4N,0)$ and the $\beta$ and $\gamma$ bands are generated by the $(4N-4,2)$ irrep with $K^\pi =0^+_\beta$ and $2^+_\gamma$ respectively. Similarly, in $sdgi$IBM these $SU(3)$ irreps are $(6N,0)$ and $(6N-4,2)$ respectively. Often we will drop $\pi$ in $K^\pi$. The quadrupole-quadrupole interaction Hamiltonian $H_Q^{(\baa)}$ is 
\be
H_Q^{(\baa)}=-(1/4)Q^2(\baa) \cdot Q^2(\baa) = -\cc_2(SU^{(\baa)}(3)) +(3/4)L^2
\label{eq.new-2}
\ee
and its eigenvalues over a $SU^{(\baa)}(3)$
state $\l|(\la \mu)KL\ran$ of $N$ bosons are $-[\la^2 +\mu^2 + \la \mu +3(\la  + \mu)] +
\f{3}{4}L(L+1)$. Therefore, all the $H_Q^{(\baa)}$'s, for a given $\eta$ value, generate the same spectrum.
Our interest here is to investigate $E2$ decay properties of the $\beta$ and $\gamma$ band levels transition to the $g$ band levels in both $sdg$IBM and $sdgi$IBM. Quadrupole moments of $g$ band levels and $g \rightarrow g$ $B(E2)$'s are studied before in 
\cite{KSP-19,Kota-book,EPJST}. To this end we will first determine the intrinsic structure of the $g$, $\beta$ and $\gamma$ bands in IBM.

\subsection{Structure of $g$, $\beta$ and $\gamma$ intrinsic states}

In the large $N$ limit, structure of the $g$, $\beta$ and $\gamma$ band levels follow from the corresponding intrinsic states and their forms, within a normalization factor, are
\be
\barr{l}
\l|N: K=0\ran_g = \l(b^\dagger_{0_g}\r)^N \l|0\ran\;,\\
\l|N: K=0\ran_\beta = \l(b^\dagger_{0_g}\r)^{N-1} b^\dagger_{0_\beta}\l|0\ran\;,\\
\l|N: K=2\ran_\gamma = \l(b^\dagger_{0_g}\r)^{N-1} b^\dagger_{2_\gamma}\l|0\ran\;.
\earr \label{eq.new-3}
\ee
The $b^\dagger_{K_o}$, with $o=g$ or $\beta$ or $\gamma$,  is a deformed single boson creation operator and it is a linear combination of the single particles creation operators $b^\dagger_{\ell , K}$ where $K$ is the $\ell_z$ eigenvalue. Thus,
\be
b^\dagger_{K_o} = \dis\sum_{\ell} x_{\ell,K}\;b^\dagger_{\ell ,K} 
\label{eq.new-4}
\ee
where $x_{\ell,K}$ are the expansion coefficients.
There are many ways to determine these coefficients; see for example 
\cite{Kota-book,KM-1}. Following \cite{KM-1}, the eigenvalue equation satisfied by $x_{\ell,K}$ is
\be
\dis\sum_{\ell_1} \lan \ell_1 K\;\;\ell_2 -K \mid 20\ran\,
t^{(\eta)}_{\ell_1,\ell_2} \,x_{\ell_1,K} = \lambda_K\, x_{\ell_2,K}\;.
\label{eq.new-5}
\ee
The $t$'s here are defined in Eq. (\ref{eq.new-1}) and $\lan --\mid--\ran$ is a Clebsch-Gordon coefficient. With $\ell=0$, 2 and 4 for $sdg$IBM, for $K=0$ we need to diagonalize a $3 \times 3$ matrix with the eigenvector for the highest eigenvalue giving $b^\dagger_{0_g}$ and the next eigenvalue gives
$b^\dagger_{0_\beta}$. Similarly, for $K=2$ we have a $2 \times 2$ matrix with the eigenvector for the highest eigenvalue  giving $b^\dagger_{2_\gamma}$. Diagonalization of the matrices given by Eq. (\ref{eq.new-5}), 
the following $g$, $\beta$ and $\gamma$ states are obtained  for a given $(\baa) = (\alpha_{02} , \alpha_{24}) = (\alpha_{sd} , \alpha_{dg})$,
\be
\barr{rcl}
b^\dagger_{0_g} & = & \dis\sqrt{\dis\f{7}{35}}\, s^\dagger + \alpha_{sd}\,\dis\sqrt{\dis\f{20}{35}}\, d^\dagger_0 + \alpha_{sd} \alpha_{dg} \, \dis\sqrt{\dis\f{8}{35}} \,g^\dagger_0 \;,\\
b^\dagger_{0_\beta} & = & \dis\sqrt{\dis\f{56}{210}}\, s^\dagger + \alpha_{sd}\,\dis\sqrt{\dis\f{10}{210}}\, d^\dagger_0  - \alpha_{sd} \alpha_{dg} \, \dis\sqrt{\dis\f{144}{210}} \, 
g^\dagger_0 \;,\\
b^\dagger_{2_\gamma} & = & \dis\sqrt{\dis\f{1}{7}} \,
d^\dagger_2 + \alpha_{dg} \, \dis\sqrt{\dis\f{6}{7}} \,
g^\dagger_2 \;.
\earr \label{eq.new-6}
\ee
Note that for $sdg$IBM, $(\alpha_{sd} , \alpha_{dg})=(+,+)$, $(+,-)$, $(-,+)$ and $(-,-)$. The result in Eq. (\ref{eq.new-6}) for the $g$ state is known before \cite{KSP-19} and the other two are new. It is remarkable that they have simple dependence on $\alpha_{\ell , \ell+2}$. 

Applying Eq. (\ref{eq.new-5}), we have in $sdgi$IBM for $K=0$ a $4 \times 4$ matrix and for $K=2$ a $3 \times 3$ matrix. Solving these the following $g$, $\beta$ and $\gamma$ states are obtained  for a given $(\baa) = (\alpha_{02} , \alpha_{24}, \alpha_{46}) = (\alpha_{sd} , \alpha_{dg} , \alpha_{gi})$,
\be
\barr{rcl}
b^\dagger_{0_g} & = & \dis\sqrt{\dis\f{33}{231}} \,s^\dagger + \alpha_{sd}\,\dis\sqrt{\dis\f{110}{231}}\, d^\dagger_0 + \alpha_{sd} \alpha_{dg} \, \dis\sqrt{\dis\f{72}{231}}\, g^\dagger_0 \\
& + & \alpha_{sd} \alpha_{dg} \alpha_{gi}\, \dis\sqrt{\dis\f{16}{231}}\, i^\dagger_0 \;,\\
b^\dagger_{0_\beta} & = & \dis\sqrt{\dis\f{66}{385}}\, s^\dagger + \alpha_{sd}\,\dis\sqrt{\dis\f{55}{385}}\, d^\dagger_0 - \alpha_{sd} \alpha_{dg} \, \dis\sqrt{\dis\f{64}{385}}\, g^\dagger_0 \\
& - & \alpha_{sd} \alpha_{dg} \alpha_{gi}\, \dis\sqrt{\dis\f{200}{385}}\, i^\dagger_0 \;,\\
b^\dagger_{2_\gamma}\l|0\ran & = & \,\dis\sqrt{\dis\f{11}{231}}\, d^\dagger_2 + \alpha_{dg} \, \dis\sqrt{\dis\f{108}{231}}\, g^\dagger_2 \\
& + & \alpha_{dg} \alpha_{gi}\, \dis\sqrt{\dis\f{112}{231}}\, i^\dagger_2  \;.
\earr \label{eq.new-7}
\ee
Note that for $sdgi$IBM, $(\alpha_{sd} , \alpha_{dg}, \alpha_{gi})=(+,+,+)$, $(+,+,-)$, $(+,-,+)$ and $(+,-,-)$,
$(-,+,+)$, $(-,+,-)$, $(-,-,+)$ and $(-,-,-)$. The result in Eq. (\ref{eq.new-7}) for the $g$ state is known before \cite{EPJST} and the other two are new. It is remarkable that they have simple dependence on $\alpha_{\ell , \ell+2}$ just as in $sdg$IBM.
The $g \rightarrow g$, $\beta \rightarrow g$ and $\gamma \rightarrow g$ $B(E2)$'s 
are determined by $t^{(\eta)}_{\ell_1 , \ell_2}$ given by Eq. (\ref{eq.new-1}) 
and the $x_{\ell ,K}$ given by Eqs. (\ref{eq.new-6}) and (\ref{eq.new-7}). We will now turn to these.

\subsection{Results for $B(E2)$ values}

Carrying out angular momentum projection from the intrinsic states given by Eq. (\ref{eq.new-3}), formulas are derived in \cite{KM-1} for $B(E2)$'s that are good for sufficiently large values of $N$. In all the calculations, the $E2$ transition operator is chosen to be \cite{KSP-19,EPJST},  
\be
T^{E2} = q_2\,Q^2_q(+,+,\ldots)\;b^2
\label{eq.new-8}
\ee
where $q_2$ is a parameter and $b$ is the oscillator length constant. Note that in the $E2$ operator all $\alpha_{\ell , \ell+2}=+1$. The form chosen in Eq. (\ref{eq.new-8}) is standard in IBM literature \cite{Iac} [it will not change with $(\baa)$] and the $sd$IBM example described in Appendix-A further confirm this is appropriate. Normally the choice of the phases in the transition operator vis-a-vis the phase choice in the Hamiltonian is done in accordance to the most relevant reproduction of transition rates etc. or by imposing other specific physical criteria. It is possible that in different parts of the periodic table or in different parts of the spectrum of a given nucleus, the phases choices in the effective nuclear Hamiltonians may 
vary and hence the interest to examine different combinations of phases. Data analysis that will be considered in a furture publication is expected to shed more light on this important issue; see also Section 6.

Firstly, formula for $B(E2)$'s is,
\be
B(E2;L_i \rightarrow L_f)= \dis\frac{5}{16\pi}\;\dis\f{\l|\lan L_f \mid\mid T^{E2}
\mid\mid L_i\ran\r|^2}{(2L_i+1)} \;.
\label{eq.new-9}
\ee
For $g \rightarrow g$ and $\beta \rightarrow g$ transitions the reduced matrix elements of the $Q^2(\baa)$ are given by \cite{KM-1},
\be 
\barr{l}
\lan N;K=0,(L+2)_f \mid\mid Q^2(+,+,\ldots) \mid\mid N;K=0,L_g\ran =
\l[N\dis\sqrt{(2L+1)}\r] \lan L 0\;\;20 \mid L+2,0\ran \\
\times\;\l[B_{00} +\frac{1}{N}\l(B^X_{00} - \dis\frac{B_{10}-3B_{00}}{a}\r) 
-\dis\frac{L(L+3)}{aN^2}\l\{B_{00} - \dis\frac{F}{12a}
\r\}\r]\;;\\
B_{mn}=\dis\sum_{\ell^\pr , \ell}\;\l[\ell^\pr(\ell^\pr+1)\r]^m 
\l[\ell (\ell+1)\r]^n \lan \ell^\pr
0\;\ell 0 \mid 20\ran \; t^{(\eta)}_{\ell^\pr,\ell}\;x_{\ell^\pr ,0} x_{\ell ,0}\;, \\
F=B_{20}-B_{11} + 6B_{10}-12B_{00}, \\
a=\dis\sum_{\ell} \ell (\ell+1) \l(x_{\ell ,0}\r)^2 \;.
\earr \label{eq.new-10}
\ee
Here, the $x_{\ell ,0}$ are the expansion coefficients for the $g$ state given by Eqs. (\ref{eq.new-6}) and (\ref{eq.new-7}). Also, for $g \rightarrow g$ transitions $B^X_{00}=B_{00}$ and for $\beta \rightarrow g$ transitions $B^X_{00}=B_{00}$ with $x_{\ell ,0}$ being the expansion coefficients for the $\beta$ state given in Eqs. (\ref{eq.new-6}) and (\ref{eq.new-7}). It is important to mention that the $t^{(\eta)}$ in Eq. (\ref{eq.new-10}) follow from Eq. (\ref{eq.new-1}) with $\alpha_{\ell , \ell+2}=+1$ for all $\ell$ values. Going further, the $B(E2)$'s for $\gamma \rightarrow g$ transitions are given by Eq. (\ref{eq.new-9}) along with \cite{KM-1}, 
\be
\barr{l}
\lan N;K=2_\gamma,(L+2)_\gamma \mid\mid Q^2(+,+,\ldots) \mid\mid N;K=0,L_g\ran \\
=  \dis\sqrt{2N\;(2L+1)} \lan L0\;\;22 \mid L+2,2\ran \;\dis\sum_{\ell^\pr , \ell}\,
t^{(\eta)}_{\ell^\pr,\ell}\;x_{\ell^\pr ,2} x_{\ell ,0} \lan \ell^\pr 2\;\;\ell 0\mid 2 2\ran\;.
\earr \label{eq.new-11}
\ee
The $x_{\ell ,0}$ and $x_{\ell ,2}$ are the expansion coefficients for the $g$ and $\gamma$ states respectively as given by Eqs. (\ref{eq.new-6}) and (\ref{eq.new-7}). Again as above, the $t^{(\eta)}$ in Eq. (\ref{eq.new-11}) follow from Eq. (\ref{eq.new-1}) with $\alpha_{\ell , \ell+2}=+1$ for all $\ell$ values. 
Using Eqs. (\ref{eq.new-10}) and (\ref{eq.new-11}), in $sdg$IBM and $sdgi$IBM we have calculated $B(E2; L_\beta \rightarrow (L-2)_g)$ and $B(E2; L_\gamma \rightarrow (L-2)_g)$ for $L=2$, $4$ and $6$ and also $B(E2)$'s for the $g$ band levels for all choices of $\baa$. In all these calculations, the $T^{E2}$ operator is given by  Eq. (\ref{eq.new-8}). The results are given in Tables 1 and 2. Note that the results for the $g$ band members with $L=2$ and $4$ are given before in \cite{KSP-19,Kota-book,EPJST}. 

Results in Table 1 for $sdg$IBM show that firstly for $(\alpha_{sd},\alpha_{dg})=(+,+)$, the $E2$ transitions from $\beta$ and $\gamma$ bands to $g$ bands are forbidden as the $Q$ in $H_Q$ and the $E2$ operator are same. However, for other choices of $(\alpha_{sd},\alpha_{dg})$ this will not apply and hence they give new results. As seen from Table 1,  for $(\alpha_{sd},\alpha_{dg})=(-,+)$  the $\beta \rightarrow g$ transition strengths are strong compared to $g \rightarrow g$ and $\gamma \rightarrow g$ strengths. However, for $(\alpha_{sd},\alpha_{dg})=(-,-)$  the $\beta \rightarrow g$ transition strengths are much weaker compared to $\gamma \rightarrow g$ strengths. All three are of similar strength for $(\alpha_{sd},\alpha_{dg})=(+,-)$. 

Results in Table 2 for $sdgi$IBM show that firstly for $(\alpha_{sd},\alpha_{dg}, \alpha_{gi})=(+,+,+)$ the $E2$ transitions from $\beta$ and $\gamma$ bands to $g$ bands are forbidden as the $Q$ in $H_Q$ and the $E2$ operator are same. However, for other choices of $(\alpha_{sd},\alpha_{dg})$ this will not apply. It is seen from the table that the $\beta \rightarrow g$ transitions are very weak compared to $g \rightarrow g$ and $\gamma \rightarrow g$ transitions for $(\baa) = (+,-,+)$ and $(-,-,-)$. Similarly, the $\gamma \rightarrow g$ transitions are very weak compared to the other two for $(\baa) =(-,+,+)$. For the rest of the choices of ($\baa$), the $E2$ transitions structure is mixed.   
These combined with the conclusions from Table 1 and the results reported before in \cite{KSP-19,EPJST} allow one to identify two generic results: (i) with multiple $SU(3)$ algebras, it is possible to have rotational bands with very weak $E2$ strengths among the levels where normally one expects strong strengths; (ii) $E2$ decay of the levels of $\beta$ and $\gamma$ bands to the ground band are quite different for some of the $SU^{(\baa)}(3)$ algebras with strong dependence on ($\baa$).  
\begin{table}
\caption{$B(E2; L \rightarrow L-2)$ values for $g \rightarrow g$, $\beta \rightarrow g$ and $\gamma \rightarrow g$ transitions for a 10 boson system generated by the four $SU^{(\baa)}(3)$ algebras in  $sdg$IBM. The $B(E2; L \rightarrow L-2)$ (in units of $(q_2)^2\,b^2)$ are given for $L=2$, $4$ and $6$.}
\label{table1}
\begin{tabular}{cccccc}
\hline
$L$ & $\;\;\;\;\;$ & \multicolumn{4}{c}{$B(E2; L \rightarrow L-2)$}  \\
\hline
& & $(+,+)$ & $(-,+)$ & $(+,-)$ & $(-,-)$ \\
\hline
$2$ & $g \rightarrow g$ & $137.05$ &  $3.16$ & $27.46$ & $21.99$ \\
& $\beta \rightarrow g$ & $0$ & $42.44$ & $22.91$ &  $2.99$ \\
& $\gamma \rightarrow g$ & $0$ & $10.19 $ & $39.1$ & $89.2$ \\
\hline
$4$ & $g \rightarrow g$ & $194.61$ & $4.77$ & $37.84$ & $31.54$ \\
& $\beta \rightarrow g$ & $0$ & $60.63$ & $32.73$ & $4.27$ \\
& $\gamma \rightarrow g$ & $0$ &  $6.06$ & $23.27$ & $53.09$ \\
\hline
$6$ & $g \rightarrow g$ & $212.02$ & $5.77$ & $38.98$ & $34.98$ \\
& $\beta \rightarrow g$ & $0$ & $66.78$ & $36.05$ &  $4.70$ \\
& $\gamma \rightarrow g$ & $0 $ & $4.99$ & $19.14$ & $43.66$ \\
\hline
\end{tabular}
\end{table}
\begin{table}
\caption{$B(E2; L \rightarrow L-2)$ values for $g \rightarrow g$, $\beta \rightarrow g$ and $\gamma \rightarrow g$ transitions for a 10 boson system generated by the eight $SU^{(\baa)}(3)$ algebras in  $sdgi$IBM. The $B(E2; L \rightarrow L-2)$ (in units of $(q_2)^2$\, $b^2$) are given for $L=2$, $4$ and $6$.}
\label{table2}
\begin{tabular}{cccccccccc}
\hline
$L$ & $\;\;\;\;\;$ & \multicolumn{8}{c}{$B(E2; L \rightarrow L-2)$}  \\
\hline
& & $(+,+,+)$ & $(+,+,-)$ & $(+,-,+)$ & $(+,-,-)$ & $(-,+,+)$ & $(-,+,-)$ & $(-,-,+)$ & $(-,-,-)$ \\
\hline
$2$ & $g \rightarrow g$ & $902.95$ & $637.54$ & $112.04$ & $33.47$ & $135.07$ & $46.54$ & $61.51$ & $159.82$ \\
& $\beta \rightarrow g$ & $0$ & $92.26$ & $3.05$ & $128.89$ & $189.43$ & $17.29$ & $144.37$ & $5.81$ \\
& $\gamma \rightarrow g$ & $0$ & $49.24$ & $130.33$ & $339.78$ & 
$14.03$ & $115.84$ & $229.88$ & $491.91$ \\
\hline
$4$ & $g \rightarrow g$ & $1286.43$ & $901.57$ & $158.16$ &  $45.36$ & $195.47$ & $66.27$ & $86.67$ & $229.54$ \\
& $\beta \rightarrow g$ & $0$ & $131.80$ & $4.36$ & $184.13$ &$270.61$ & $24.70$ & $206.25$ & $8.30$ \\
& $\gamma \rightarrow g$ & $0$ & $29.31$ & $77.57$ & $202.25$ & $8.35$ & $68.95$ & $136.84$ & $292.80$ \\
\hline
$6$ & $g \rightarrow g$ & $1409.96$ & $974.87$ & $170.47$ &   $45.29$ & $220.33$ & $72.54$ & $93.12$ & $255.24$ \\
& $\beta \rightarrow g$ & $0$ & $145.16$ & $4.81$ & $202.80$ & $298.05$ & $27.20$ & $227.16$ & $9.14$ \\
& $\gamma \rightarrow g$ & $0$ & $24.10$ & $63.80$ & $166.33$ & $6.87$ & $56.71$ & $112.53$ & $240.79$ \\
\hline
\end{tabular}
\end{table}
  
\section{Properties of $\gamma$ ($K=2$) band in shell model: DSM results for a $(sdg)^{8p}$ system}

In our previous work \cite{KSP-19,EPJST} we have carried out SM and DSM studies in $sdg$ and $sdgi$ spaces for systems that give the lowest $SU(3)$ irrep to be $(\lambda, 0)$ type with $\lambda$ even. Then, we have uniquely a $g$ band with  $L=J=0$, $2$, $\ldots$, $\lambda$. This is true for nucleon (fermion) numbers $2$, $6$, $12$ etc with spin $S=0$. However, the next lowest irrep often for these do not give a $K=2$ (i.e. $\gamma$) band with $S=0$; see \cite{Ko-irreps} and also Tables 3.1-3.3 in \cite{Kota-book}. Considering nucleons numbers $4$, $8$, $10$ and so on, one sees that the lowest $SU(3)$ irrep with $S=0$ for these will be $(\lambda , \mu)$ with $\lambda$ even and $\mu=2$ or $4$ or $6$ etc. Then, (for $\mu < \lambda$), the $SU(3)$ irrep $(\lambda , \mu)$ generates $K=0$, $2$, $\ldots$, $\mu$ bands. Thus, in this situation, for fermion systems, the $g$ band and the $\gamma$ band are generated by the same $SU(3)$ irrep. 
Applying the methods described in \cite{Ko-irreps,Kota-book} for $m$ number of protons ($p$) or neutrons ($n$) in $(sdg)$ orbits (or $\eta =4$ shell) with $m=4$, $8$, $10$ and $14$ and similarly in $(sdgi)$ orbits (or $\eta=6$ shell) with $m=4$, $8$, $10$, $14$ and $18$, it is easy to see that the leading or highest weight $SU(3)$ irreps with $S=0$ (then J=L) are,
\be
\barr{l}
(sdg)^4 \;:\; (12,2),\;\;\;(sdg)^8 \;:\; (18,4),\;\;\;(sdg)^{10} \;:\; (20,4),\;\;\;(sdg)^{14} \;:\; (20,6), \\
(sdgi)^4\;:\; (20,2),\;\;\;(sdgi)^8 \;:\; (34,4),\;\;\;(sdgi)^{10} \;:\; (40,4),\\
(sdgi)^{14} \;:\; (48,6),\;\;\;(sdgi)^{18} \;:\; (54,6)\;. \\
\earr \label{eq.new-12}
\ee
Then, as a first step we need to understand the properties of the $g$ ($K=0$), $\gamma$ ($K=2$) and other higher bands generated by the above $SU(3)$ irreps say in $(sdg)$ and $(sdgi)$ spaces by multiple $SU(3)$ algebras. For example, $(sdg)^{8p}$ gives $(18,4)$ irrep generating $K=0$ ($L=0$, 2,4, $\ldots$), $K=2$ ($L=2$, 3, 4, $\ldots$) and $K=4$ ($L=4$, 5, 6, $\ldots$)  bands. 
Choosing $H_Q^{(\baa)}$ as in Eq. (\ref{eq.new-2}) but for fermions, its eigenvalues clearly do not depend on $K$. Therefore, the $(18,4)$ irrep gives
two degenerate $2^+$ levels, three degenerate $4^+$ levels, two degenerate $5^+$ levels, three degenerate $6^+$ levels etc. In order to remove the degeneracies in the spectrum  and also for proper spacing between the band heads of the $K=0$, $2$ and $4$ bands, it is necessary to include the $SU(3) \supset SO(3)$ integrity basis operators that are $3-$ and $4-$body interactions \cite{Draayer1,Draayer2}. These will mix a predefined $K$ (examples are the $K$'s defined in \cite{El58-2,Verg,AkiDra}) but not the $(\lambda \mu)$ and $L$. The integrity basis operators involve $Q^2_\mu(\baa)$ and therefore they carry not only the effects due to their $3$ and $4$-body character (they produce many new effects as described for example in \cite{Bon-1,Bon-2,Bon-3,Bon-4} that are not possible with only two-body $SO(3)$ scalars in $SU(3)$) but also due to $(\baa)$. Thus, the analysis with these operators is more complex.
Developing SM or DSM codes including these $3-$ and $4-$ body interactions is not straight forward and therefore the study of $\gamma$ and higher $K$ bands in SM using higher-body interactions is postponed to future. Here below, we will present some results obtained using DSM (without higher-body interactions) for the $\gamma$ band with multiple $SU(3)$ algebras by using some additional constraints as described below.

\subsection{DSM results for $\gamma$ band for a $(sdg)^{8p}$ system}

\begin{figure}
\includegraphics[width=0.5\linewidth]{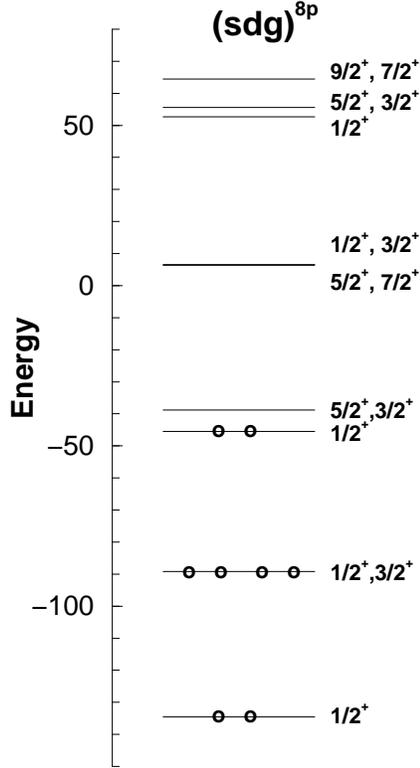}
\caption{Hartree-Fock sp spectrum corresponding to the lowest intrinsic state for a $(sdg)^{8p}$ system generated by the four $H_Q$ operators	in Eq. (\ref{eq.new-2}). 
In the figure, the symbol $\mbox{o}$ denotes protons. 
Shown in the figure are the $k$ values of the sp orbits and
each orbit is doubly degenerate with $\l|k\ran$ and $\l|-k\ran$ states. The
spectra  are same for all  the four Hamiltonians although the sp wavefunctions are different. Note that the energies in the figures are unitless and the unit MeV has to be put back after multiplying with an appropriate scale factor if the results are used for a real nucleus. See text for further details.}
\label{fig-new}
\end{figure}

In our DSM study of $\gamma$ band, considered is the example
of 8 protons in the $sdg$ shell. The Hamiltonian used is same as 
the $H^{(\baa)}_Q$ in Eq. (\ref{eq.new-2}) with the $(b^\dagger_\ell \tilde{b}_{\ell^\prime})^2_\mu$ in the $Q^2(\baa)$ operator in Eq. (1) is replaced by 
$$
\dis\sqrt{2}\;(a^\dagger_{\ell \f{1}{2}} \tilde{b}_{\ell^\prime \f{1}{2}})^{2,0}_\mu \;.
$$ 
Note that $a^\dagger$ and $a$ are fermion (in our example protons) creation and annihilation operators with $\tilde{a}_{\ell m,\f{1}{2}\, m_s} = (-1)^{\ell-m+\f{1}[2] -m_s}
a_{\ell\, -m,\f{1}{2} -m_s}$. Similarly, $\ell$ is orbital angular momentum and $\f{1}{2}$ is spin.
With ($sdg$) orbits, we have $(\baa) = 
(\alpha_{sd} , \alpha_{dg})$ and it takes four values. For all the 
four $H_Q$'s DSM calculations are performed. The HF sp spectrum obtained 
with the corresponding lowest HF intrinsic state
is presented in Fig. \ref{fig-new}. The HF sp energies do not depend on $\baa$ 
but the sp wavefunctions are different (see for example Table 3 in \cite{EPJST}). By exciting two valence protons to the
orbits $k=5/2^+_1$ and/or $k=3/2^+_2$, and then performing a self-consistent HF
calculation for the rest of the particles (tagged HF) \cite{tagg1,tagg2,tagg3},
we have generated two excited $K=0$ intrinsic
states, two $K=2$ and two $K=4$ intrinsic states. 
The two excited intrinsic states with $K=0$ have the structure 
$(1/2)_1^2(1/2)_2^2(3/2)_1^2(3/2)_2^2$ and 
$(1/2)_1^2(1/2)_2^2(3/2)_1^2(5/2)_1^2$. Similarly the
two intrinsic states with $K=2$ have the structure $(1/2)_1^2(1/2)_2^2
(3/2)_1^2(3/2)_2^1(1/2)_3^1$ and 
$(1/2)_1^2(1/2)_2^2(3/2)_1^2(5/2)_1^1(1/2)_3^1$. 
The two $K=4$ intrinsic states have the structure $(1/2)_1^2(1/2)_2^2
(3/2)_1^1(5/2)_1^1(1/2)_3^2$ and 
$(1/2)_1^2(1/2)_2^2(3/2)_1^2(5/2)_1^1(3/2)_2^1$. 
In the above, the superscript gives the number of protons and subscript gives the serial number of the sp level with a given $k$. Using these six intrinsic states plus the lowest intrinsic state shown in Fig. 1 (total seven intrinsic states), angular momentum 
projection and band mixing calculations are carried out. 
We found that to reproduce the degeneracy of the $2^+$ and $4^+$ levels, as expected from $SU(3)$ [see Eq. (\ref{eq.new-12})],
we have to multiply the $Q \cdot Q$ two-body matrix elements 
by a scaling factor 0.97. The scaling factor simulates the effect of the $SU(3) \supset SO(3)$ integrity basis operators that are 3 and 4-body operators and the analysis in \cite{DrRo}, where correlation coefficients and norms of operators are used, supports this. Not only all the four $H_Q$'s generate the same spectrum but more importantly, as expected from the lowest irrep $(18,4)$ of the four $SU^{(\baa)}(3)$ algebras, we have $K=0$ ($g$), $K=2$
($\gamma$) and $K=4$ bands [see Eq. (\ref{eq.new-12})] with
two nearly degenerate $2^+$ states at
$\sim 4.5$ excitation from the ground state, three $4^+$ states at $\sim 15$  and three $6^+$ states at $\sim 31.5$ (deviations being less than $5\%$). 

Going further,
$B(E2)$ values are calculated using the $E2$ operator 
$$
T^{E2}=e_p\;Q^2_q(-,-)\,b
$$ 
with the the oscillator length parameter $b=1\,fm$ and $e_p=1e$. 
The choice of using $Q^2_q(\baa)$ with $(\baa) = (-,-,\dots)$ is made in all shell model (and DSM) studies; see \cite{SM-MM1,SM-MM2,KS-book} and Appendix-A. 
Results for $B(E2)$'s involving lowest $2^+$ and $0^+$ states are given in Table \ref{tab-new}. It is easy to see that the $g \rightarrow g$ transitions are strong for $(\baa)=(-,-)$ as here the $E2$ operator chosen is same as the $Q$ operator in $H_Q$. The $B(E2)$ are also strong for $(+,-)$ choice. Results in Table \ref{tab-new} show that the $(-,+)$ system gives $g \rightarrow g$ and $\gamma \rightarrow g$ transition strengths comparable. However, the $\gamma \rightarrow g$ transitions quite weak compared to $g \rightarrow g$ transitions for the other three $(\baa)$ choices showing good band structure (see also the results for $4^+$ of $K=4$ band). Thus, the $\gamma$ band structure depends on $(\baa)$. A more detailed DSM analysis of all the bands from the irreps listed in Eq. (\ref{eq.new-12}) is postponed to a future investigation.

\begin{table}
\caption{DSM results, given in columns 2-5, for some $B(E2)$ values (in $e^2\;fm^4$ units) for a $(sdg)^{8p}$ system }
\label{tab-new}
\begin{tabular}{ccccc}
\hline
$J_f \rightarrow J_i$ & $(-,-)$ & $(-,+)$ & $(+,-)$ & $(+,+)$ \\
\hline
$2^+_g \rightarrow 0^+_g$       &   747  &   18      &   303  &    32\\
$2^+_\gamma \rightarrow 0^+_g$  &    25  &   27   &   0.2    &   0.4\\
$4^+_g \rightarrow 2^+_g$       &   1052 &   30   &   414    &   43\\
$4^+_\gamma \rightarrow 2^+_g$  &     7  &   11.3 &   0.1    &   1.2\\
$4^+_\gamma \rightarrow 2^+_\gamma$ &427 &    5.3 &   181    &    24\\
$4^+_{K=4} \rightarrow 2^+_g$   &   0    &     0  &   0.01   &  0.01\\
$4^+_{ K=4 } \rightarrow 2^+_\gamma$&26    &    24  &   1.1    &  0.7\\
\hline
\end{tabular}
\end{table}

\begin{figure}
\includegraphics[width=0.6\linewidth]{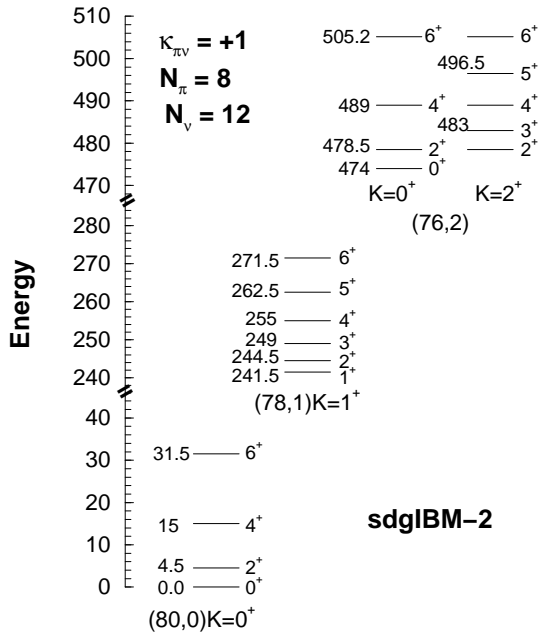}
\caption{Rotational bands from the lowest three $SU^{(\baa_\pi , \baa_\nu)}(3)$ irreps for a $N_\pi=8$ and $N_\nu=12$ system in $sdg$IBM-2 with $\kappa_{\pi \nu}=+1$. For each band, levels only up to $J^\pi=6^+$ are shown. It is important to note that the spectrum is independent of $(\baa_\pi , \baa_\nu)$ values, i.e. the spectrum is same for all the 16 $SU^{(\baa_\pi , \baa_\nu)}(3)$ algebras. The scissors $K=1^+$ band in the figure can be pushed up in energy by including a Majorana term in the Hamiltonian in Eq. (\ref{eq.new-15}) \cite{Iac}. The spectrum is clearly close to a axially symmetric deformed nucleus. In the figure, the energies are unitless and the unit MeV has to be put back after multiplying with an appropriate scale factor if the results are used for a real nucleus.} 
\label{fig-1}  
\end{figure}

\section{Multiple $SU(3)$ algebras in IBM-2: Results for scissors $1^+$ band in $sdg$IBM-2 and $sdgi$IBM-2}

\subsection{Introduction to multiple $SU(3)$ algebras in
IBM-2}

In proton-neutron IBM or IBM-2, $SU(3)$ algebra appears both in proton ($\pi$) and neutron ($\nu$) spaces. Therefore, in the proton-neutron space (i.e. in the product space), there will be much larger number of $SU(3)$ algebras. In $sd$IBM-2, some of the multiple $SU(3)$ algebras are analyzed in \cite{Scholten}. Going beyond $sd$IBM-2, in the first non-trivial $sdg$IBM-2 there are four $SU(3)$ algebras in proton space generated by $[Q^2_\pi(\baa_\pi) , L_\pi]$ operators and four in neutron space generated by $[Q^2_\nu(\baa_\nu) , L_\nu]$ operators. As the total quadrupole operator is,
\be
Q^2(\baa_\pi , \baa_\nu : \kappa_{\pi \nu}) = Q^2_\pi(\baa_\pi) + \kappa_{\pi \nu} Q^2_\nu(\baa_\nu)\;;\;\;\;\kappa_{\pi \nu}=\pm 1\;,
\label{eq.new-13}
\ee
we have a total of 16 $SU(3)$ algebras in $sdg$IBM-2 for each $\kappa_{\pi \nu}$. Note that, the $Q^2_\rho(\baa_\rho)$ with $\rho=\pi,\nu$ are defined by Eq. (\ref{eq.new-1}) but in $\pi$ and $\nu$ spaces respectively. In $sdgi$IBM, we will have 64 $SU(3)$ algebras for each $\kappa_{\pi \nu}$ value. In general for proton and neutron bosons in a $\eta$ shell ($\eta=2$ for $sd$, $\eta=3$ for $pf$, $\eta=4$ for $sdg$ and so on), the number of $SU(3)$ algebras is $2^{\l[\f{\eta}{2}\r]} \times 2^{\l[\f{\eta}{2}\r]}$ for each $\kappa_{\pi \nu}$ value. In this Section we will consider multiple $SU(3)$ algebras in $sdg$IBM-2 and $sdgi$IBM-2.  

Given number of proton bosons $N_\pi$ and neutron bosons $N_\nu$, the total boson number is $N=N_\pi + N_\nu$. Then, the $g$ bands in $\pi$ and $\nu$ spaces are generated by $(\eta{N_\pi},0)$ and $(\eta{N_\nu},0)$ irreps of $SU(3)$ respectively for a $\eta$ shell ($\eta=4$ for $sdg$IBM and 6 for $sdgi$IBM). By restricting to these bands (other extensions will be considered in future publications), we have the basis states $\phi(L_\pi , L_\nu , L)$ where
\be
\phi(L_\pi , L_\nu , L) =
\l| (\eta{N_\pi},0)L_\pi , (\eta{N_\nu},0)L_\nu ; L,M\ran
\label{eq.new-14}
\ee
With $\eta$ even, we have $L_\pi = 0,2,4,\ldots,\eta{N_\pi}$,
$L_\nu = 0,2,4,\ldots,\eta{N_\nu}$ and $L = 0,1,2,\ldots \eta{(N_\pi + N_\nu)}$ without counting multiplicities. Now, the simple quadrupole-quadrupole Hamiltonian $H_Q$,
\be
\barr{rcl}
H^{(\baa_\pi , \baa_\nu)}_Q & = & H_{Q:\pi} + H_{Q: \nu} + \kappa_{\pi \nu} H_{Q: \pi\nu}\;;\\
H_{Q:\pi} & = & -\f{1}{4} Q^2_\pi(\baa_\pi) \cdot Q^2_\pi(\baa_\pi)\;,\;\;\;H_{Q:\nu} =-\f{1}{4} Q^2_\nu(\baa_\nu) \cdot Q^2_\nu(\baa_\nu)\;,\\
H_{Q:\pi\nu} & = & -\f{1}{2} Q^2_\pi(\baa_\pi) \cdot Q^2_\nu(\baa_\nu)\;,\;\;\;\kappa_{\pi \nu}=\pm 1\;.
\earr \label{eq.new-15}
\ee
generates for example the 16 $SU^{(\baa_\pi , \baa_\nu)}(3)$ algebras in the $sdg$IBM-2 space for $\kappa_{\pi \nu}=+1$ and 16 for $\kappa_{\pi \nu}=-1$. Restricting to $g$ bands in the $\pi$ and $\nu$ spaces, the $SU(3)$ irreps in the proton-neutron spaces for $\kappa_{\pi \nu}=+1$ are given by, 
\be
(\la_1,0) \times (\la_2,0) = \dis\sum_{r=0}^{\la_2}\;(\la_1+\la_2-2r,r)\;.
\label{eq.new-16a}
\ee
Similarly, for $\kappa_{\pi \nu}=-1$ we have
\be
(\la_1,0) \times (\la_2,0)^* = \dis\sum_{r=0}^{\la_2}\;(\la_1
-r, \la_2-r)\;.
\label{eq.new-16b}
\ee
Note that, without loss of generality, we are assuming $\la_2 \leq \la_1$ in Eqs. (\ref{eq.new-16a}) and (\ref{eq.new-16b}). Also, $(\la_2 ,0)^* = (0, \la_2)$. Now, it is easy to see that the states
$$
\l|(\eta{N_\pi},0)(\eta{N_\nu},0)(\la,\mu)KL\ran
$$
with $(\la, \mu)$ given by Eq. (\ref{eq.new-16a}) are the eigenstates of $H^{(\baa_\pi , \baa_\nu)}_Q$ with $\kappa_{\pi \nu}=+1$ . Similarly, 
$$
\l|(\eta{N_\pi},0)(0,\eta{N_\nu})(\la,\mu)KL\ran\;\;\;\mbox{or}\;\;\;
\l|(0,\eta{N_\pi})(\eta{N_\nu},0)(\la,\mu)KL\ran
$$
with $(\la , \mu)$ given by Eq. (\ref{eq.new-16b}) are the eigenstates of $H^{(\baa_\pi , \baa_\nu)}_Q$ with  $\kappa_{\pi \nu}=-1$. For example, for $N_\pi=8$ and $N_\nu=12$, we have in $sdg$IBM for $\kappa_{\pi \nu}=+1$ 
$$
(\la, \mu)_K =  (80,0)_0 \oplus (78,1)_1 \oplus (76,2)_{0,2} \oplus (74,3)_{1,3} \oplus (72,4)_{0,2,4} \oplus \ldots
$$
Clearly, the $g$ ($K=0^+$) band is generated by the irrep $(\eta{N_\pi}+\eta{N_\nu},0)$ and the scissors $1^+$ band by the irrep
$(\eta{N_\pi}+\eta{N_\nu}-2,1)$. Here after, $1^+_S$ denotes the scissors $1^+$ band. The energy eigenvalues of $H_Q$ in Eq. (\ref{eq.new-15}) are given by 
\be
E((\la_\pi ,0)(\la_\nu ,0)(\la ,\mu)KL) = -[\la^2 + \mu^2 + \la \mu +3(\la + \mu)] + \f{3}{4} L(L+1)
\label{eq.new-17}
\ee
as $H^{(\baa_\pi , \baa_\nu)}_Q = -{\cal C}_2(SU^{(\baa_\pi , \baa_\nu)}(3))$ + $(3/4) L \cdot L$ independent of $\baa_\pi$ and $\baa_\nu$.  
Note that $L=0,2,4,\ldots$ for $K=0$ and $L=K,K+1,K+2,\ldots$ for $K \ne 0$. 

Eigenstates $\l|(\eta N_\pi,0)(\eta N_\nu,0)(\la,\mu)KL\ran$ can be written in terms of the basis states $\phi(L_\pi , L_\nu , L)$ by constructing the $H^{(\baa_\pi , \baa_\nu)}_Q$ Hamiltonian matrix in this basis for each $L$ value and diagonalizing. 
Construction of the $H$ matrices is simple as $H_{Q:\pi}$ and $H_{Q:\nu}$ contribute only to the diagonal matrix elements that follow from Eq. (\ref{eq.new-17}) in $\pi$ and $\nu$ spaces respectively. For $H_{Q:\pi \nu}$ the matrix elements are given by
\be
\barr{l} 
\lan L^f_\pi , L^f_\nu , L \mid -\f{1}{2} Q^2_\pi(\baa_\pi) \cdot Q^2_\nu(\baa_\nu) \mid L^i_\pi , L^i_\nu , L \ran 
= -\dis\frac{(-1)^L}{2} \l\{\barr{ccc} L & L^f_\nu & L^f_\pi \\
2 & L^i_\pi & L^i_\nu \earr \r\}\; \\
\times \;\lan L^f_\pi \mid\mid Q^2_\pi(\baa_\pi) \mid\mid L^i_\pi\ran\;\lan L^f_\nu \mid\mid Q^2_\nu(\baa_\nu) \mid\mid L^i_\nu\ran\;.
\earr 
\label{eq.new-18}
\ee
For calculating the above reduced matrix elements in $\pi$ and $\nu$ spaces used is Eq.(\ref{eq.new-10}) for the $L_g \rightarrow L_g \pm 2$ reduced matrix elements of the $Q^2$ operator in $\pi$ and $\nu$ spaces respectively. Similarly, for $L_g \rightarrow L_g $ matrix elements the formula is\cite{KM-1,EPJST},
\be
\barr{l}
\lan N_{\rho};K=0,L \mid\mid Q^2_\rho \mid\mid N_{\rho};K=0,L\ran = \l[N_\rho \dis\sqrt{(2L+1)}\r] \lan L 0\;\;20 \mid L,0\ran \; \\
\l[B_{00} +\frac{1}{N_\rho}\l(B_{00} - \dis\frac{B_{10}-3B_{00}}{a}\r)-\dis\frac{L(L+1)}{aN^2_\rho}\l\{B_{00}+\dis\frac{F_1}{4a}\r\}\r]\;;\\
F_1=B_{20}-B_{11}-10B_{10}+12B_{00}\;,\;\;\;\rho=\pi\;\;\mbox{or}\;\;\nu\;.
\earr \label{eq.new-19}
\ee
Note that $B_{mn}$ and $a$ are defined in Eq. (\ref{eq.new-10}).
It is important to mention that Eqs. (\ref{eq.new-10}) and (\ref{eq.new-19}) give quite accurate results (usually the error is less than a few percent) and further accuracy can be achieved, if needed, using the results given in \cite{KM-3,Barr} and their extensions. Explicit diagonalization $H^{(\baa_\pi , \baa_\nu)}_Q$ is carried out for various $L$ values in $sdg$IBM-2 and $sdgi$IBM-2. The resulting eigenvalues are used to verify that indeed the eigenvalues are as given by Eq. (\ref{eq.new-17}). Note that diagonalization gives the expansion coefficients $C^{--}_{--}$ in 
\be
\barr{l}
\l|(\eta{N_\pi},0)(\eta N_\nu,0)(\la,\mu)KL\ran = \\
\\
\dis\sum_{L_\pi , L_\nu}\;C^{(\la ,\mu)KL}_{L_\pi , L_\nu}\;
\l|(\eta{N_\pi},0)L_\pi , (\eta{N_\nu},0)L_\nu ; L\ran\;.
\earr \label{eq.new-20}
\ee
Although Eqs. (\ref{eq.new-17})-(\ref{eq.new-20}) apply to the situation with $\kappa_{\pi \nu}=+1$, they will extend in a simple manner to the situation with $\kappa_{\pi \nu}=-1$ (see Section 4.4). With the method for constructing $g$ and $1^+_S$ bands for $\pi - \nu$ systems given above, 
in the following subsection 4.2 we will present the formulation for calculating $B(E2)$'s and $B(M1)$'s involving levels from these bands. Results for $B(E2)$'s and $B(M1)$'s from multiple $SU(3)$ algebras in $sdg$IBM-2 and $sdgi$IBM-2 are presented in Sections (4.3, 4.4) and 4.5 respectively.

\begin{figure}
\includegraphics[width=0.75\linewidth]{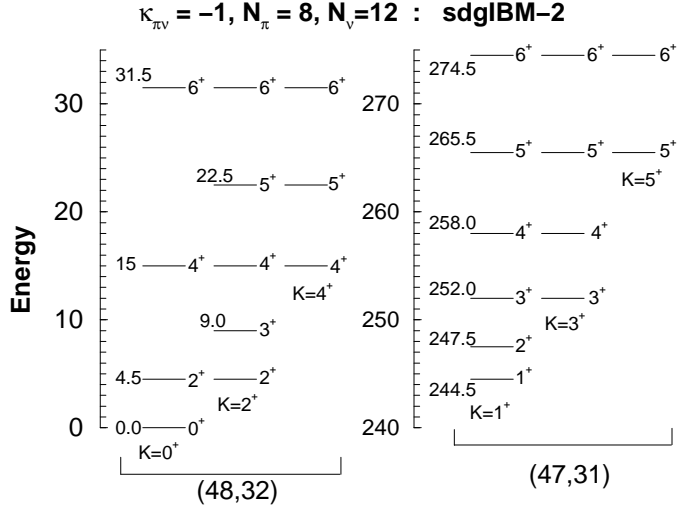}
\caption{Rotational bands from the lowest two $SU^{(\baa_\pi , \baa_\nu)}(3)$ irreps for a $N_\pi=8$ and $N_\nu=12$ system in $sdg$IBM-2 with $\kappa_{\pi \nu}=-1$.  For each band, levels only up to $J^\pi=6^+$ are shown. As $\mu$ in $(\la , \mu)$ is large here, each irrep gives large number of $K$ values (bands) and in the figure only the lowest three bands from each $(\la , \mu)$ are shown. It is clearly seen that the spectrum is close to that of a triaxial rotor. It is important to note that the spectrum is independent of $(\baa_\pi , \baa_\nu)$ values, i.e. the spectrum is same for all the 16 $SU(3)$ algebras. In the figure, energies are unitless and the unit MeV has to be put back after multiplying with an appropriate scale factor if the results are used for a real nucleus.} 
\label{fig-2}  
\end{figure}

\subsection{$M1$ and $E2$ matrix elements}

Definition of $B(E2)$'s is given  by Eq. (\ref{eq.new-9}) and for $B(M1)$'s it is 
\be
B(M1; L_i \rightarrow L_f) = \dis\f{3}{4\pi}\;\dis\f{\l| \lan L_f \mid\mid T^{M1} \mid\mid L_i\ran\r|^2}{(2L_i +1)}\;,
\label{eq.new-21}
\ee
where $T^{M1}$ is the $M1$ operator. In our calculations, we have employed 
\be
T^{M1} = \l\{g_\pi L^1_\pi + g_\nu L^1_\nu + g^B_\pi\,\l[Q^2_\pi(+,+,\ldots) \times L^1_\pi \r]^1 + g^B_\nu\,\l[Q^2_\nu(+,+,\ldots) \times L^1_\nu\r]^1 \r\}\;\mu_N
\label{eq.new-22}
\ee
for the $M1$ operator. Note that in Eq. (\ref{eq.new-22}), $\alpha_{\ell, \ell+2}=+1$ for all $\ell$ in both $\pi$ and $\nu$ spaces. In many applications in the past for example in 
$sdg$IBM-2, the above form with $g^B_\pi = g^B_\nu =0$ is employed \cite{Scholten,Piet,Kota-book}. However, the $g^B_\pi$ and $g^B_\nu$ terms are included here as it is well known that these terms are very important for proper understanding of $E2/M1$ ratios \cite{Warner,Lipas}. More importantly, without these terms, $M1$ transition strengths will be same for all $SU^{(\baa)}(3)$ algebras as $L_\pi$ and $L_\nu$ do not produce any ($\baa$) (deformation) dependence; see Tables 4 and 5. Let us add that it is possible to split the $L_\pi$ into $d$ and $g$ boson parts and similarly for the $L_\nu$ \cite{KY,Wu,KM-2} in $sdg$IBM-2 (similarly into $d$, $g$ and $i$ parts in $sdgi$IBM-2) but they are not considered in the present paper. For $B(E2)$ calculations the $E2$ operator employed is,
\be
T^{E2} = \l[q_\pi Q^2_\pi(+,+,\ldots)\; +\; q_\nu Q^2_\nu(+,+,\ldots)\r]\;b^2
\label{eq.new-23}
\ee
and this is similar to the form employed in Section 2.3 (see also Appendix-A). Note that $q_\pi$ and $q_\nu$ are proton and neutron boson effective charges. As the chosen $M1$ and $E2$ operators are of the form $T^k_\pi + U^k_\nu$, their matrix elements in the $\l|({\eta N_\pi},0)({\eta N_\nu},0)(\la,\mu)KL\ran$ states are given by
\be
\barr{l}
\lan ({\eta N_\pi},0)({\eta N_\nu},0)(\la^f,\mu^f)K^fL^f \mid\mid T^k_\pi + U^k_\nu \mid\mid ({\eta N_\pi},0)({\eta N_\nu},0)(\la^i,\mu^i)K^iL^i\ran \\
\\
= \dis\sum_{L^f_\pi L^f_\nu}\,\dis\sum_{L^i_\pi L^i_\nu}
C^{(\la^f,\mu^f)K^fL^f}_{L^f_\pi L^f_\nu}\;C^{(\la^i,\mu^i)K^iL^i}_{L^i_\pi L^i_\nu}\;(X+Y)\;;\\
X = \delta_{L^f_\nu , L^i_\nu}\;(-1)^{L^f_\pi + L^f_\nu + L^i +k}
\;\dis\sqrt{(2L^f+1)(2L^i+1)}\;\l\{\barr{ccc} L^f_\pi & L^f & L^f_\nu \\ L^i & L^i_\pi & k \earr\r\} \\
\times \lan (\eta N_\pi ,0) L^f_\pi \mid\mid T^k_\pi \mid\mid (\eta N_\pi ,0) L^i_\pi \ran\;,\\
Y = \delta_{L^f_\pi , L^i_\pi}\;(-1)^{L^i_\pi + L^i_\nu + L^f +k}
\;\dis\sqrt{(2L^f+1)(2L^i+1)}\;\l\{\barr{ccc} L^f_\nu & L^f & L^f_\pi \\ L^i & L^i_\nu & k \earr\r\} \\
\times \lan (\eta N_\nu ,0) L^f_\nu \mid\mid U^k_\nu \mid\mid (\eta N_\nu ,0) L^i_\nu \ran\;.
\earr \label{eq.new-24}
\ee
The reduced matrix elements in $X$ and $Y$ above for the $E2$ operator follow from Eqs. (\ref{eq.new-10}) and (\ref{eq.new-19}). Similarly for the $M1$ operator they are simple for the $L_\pi$ and $L_\nu$ terms. For example,
$$
\lan L^f_\pi \mid\mid L^1_\pi \mid\mid L^i_\pi \ran = \delta_{L^f_\pi\,,\,L^i_\pi}\;\sqrt{L^i_\pi (L^i_\pi+1)(2L^i_\pi +1)}\;.
$$
However, for the reduced matrix element of $\l[Q^2_\pi(+,+,\ldots) \times L^1_\pi \r]^1$ (similarly for the corresponding term with $\nu$), we have
\be
\barr{l}
\lan (\eta N_\pi ,0) L^f_\pi \mid\mid \l[Q^2_\pi(+,+,\ldots) \times L^1_\pi \r]^1 \mid\mid (\eta N_\pi ,0) L^i_\pi \ran = (-1)^{L^i_\pi + L^f_\pi +1} \; \dis\sqrt{3} \\
\times \l\{\barr{ccc} 2 & 1 & 1 \\ L^i_\pi & L^f_\pi & L^i_\pi\earr \r\}\;\sqrt{L^i_\pi (L^i_\pi+1)(2L^i_\pi +1)}\;\lan (\eta N_\pi ,0) L^f_\pi \mid\mid Q^2_\pi(+,+,\ldots) \mid\mid (\eta N_\pi ,0) L^i_\pi \ran\;.
\earr \label{eq.new-25}
\ee
Eqs. (\ref{eq.new-10}) and (\ref{eq.new-19}) will give the reduced matrix elements of $Q^2_\pi(+,+,\ldots)$. 

\begin{table}
\caption{$B(E2; L_i \rightarrow L_f)$ and $B(M1; L_i \rightarrow L_f)$ 
values for some transitions involving $g$ ($0^+$) band and $K=1^+_S$ band levels for a $N_\pi=8$ and $N_\nu=12$ system generated by the $SU^{(\baa_\pi , \baa_\nu)}(3)$ algebras with $\baa_\pi = \baa_\nu$ in $sdg$IBM-2. The $B(E2; L_i \rightarrow L_f)$ values in the table are in units of $(5/16\pi)\,b^4$. $B(M1; L_i \rightarrow L_f)$ is $\l[R(M1; L_i \rightarrow L_f)\r]^2\; (3/4\pi)\,\mu_N^2$ and given in the table are $R(M1; L_i \rightarrow L_f)$ values. Note that each $R(M1)$ is a sum of three terms given in the table. The numbers in columns 3-6 are for the four values of $\baa_\pi$ and they have to be multiplied by the factor $f$ given in the last column. In the table $1^+_S$ and $2^+_S$ are the $1^+$ and $2^+$ levels of the $1^+_S$ band. See text for more details.}
\label{table3}
\begin{tabular}{ccccccc}
\hline
$transition$ & $\;\;\;\;\;$ & $(+,+)$ & $(+,-)$ & 
$(-,+)$ & $(-,-)$ & $f$  \\
\hline
$2^+_g \rightarrow 0^+_g$ & $B(E2)$ & $13.32$ & $2.46$ & $0.41$ & $2.05$ & $\l(8\, q_\pi + 12\, q_\nu \r)^2$ \\  
$4^+_g \rightarrow 2^+_g$ & $B(E2)$ & $18.96$ & $3.5$ & $0.59$ & $2.94$ & $\l(8\, q_\pi + 12\, q_\nu \r)^2$ \\
$1^+_S \rightarrow 2^+_g$ & $B(E2)$ & $60.68$ & $13.25$ & $1.61$ & $8.35$ & $\l(q_\pi - q_\nu\r)^2$ \\ 
$2^+_S \rightarrow 2^+_g$ & $B(E2)$ & $8.47$ & $1.88$ & $0.77$ & $0.44$ & $\l(q_\pi - q_\nu\r)^2$ \\
$2^+_S \rightarrow 0^+_g$ & $B(E2)$ & $22.84$ & $3.53$ & $0.41$ & $3.96$ & $\l(q_\pi - q_\nu\r)^2$ \\
\hline
$1^+_S \rightarrow 0^+_g$ & $R(M1)$ & $3.6$ & $3.6$ & $3.6$ & $3.6$ & $(g_\nu -g_\pi)$ \\
& & $-9.52$ &$-4.65$ & $-2.44$ & $2.43$ & $8\, g^B_\pi$ \\
& & $9.38$ & $4.28$ & $1.99$ & $-3.11$ & $12\, g^B_\nu $ \\  
$1^+_S \rightarrow 2^+_g$ & $R(M1)$ & $2.59$ & $2.59$ & $2.59
$ & $2.59$ & $(g_\nu -g_\pi)$ \\
& & $-6.86$ & $-3.35$ & $-1.76$ & $1.75$ & $8\, g^B_\pi$ \\
& & $6.76$ & $3.08$ & $1.43$ & $-2.24$ & $12\,g^B_\nu$ \\  
$2^+_S \rightarrow 2^+_g$ & $R(M1)$ & $4.43$ & $4.43$ & $4.43$ & $4.43$ & $(g_\nu -g_\pi)$ \\
& & $-11.73$ & $-5.74$ & $-3.05$ & $2.94$ & $8\, g^B_\pi$ \\
& & $11.56$ & $5.29$ & $2.49$ & $-3.78$ & $12\, g^B_\nu$ \\  
\hline
\end{tabular}
\end{table}

\subsection{$B(E2)$ and $B(M1)$ results from $sdg$IBM-2: $\kappa_{\pi \nu}=+1$}

Low-lying rotational bands for a $sdg$IBM-2 system with $N_\pi=8$ and $N_\nu=12$ are shown in Fig. 2. These correspond to $(\la , \mu) =(80,0)$, $(78,1)$ and $(76,2)$ bands.
Firstly, the $H^{(\baa_\pi , \baa_\nu)}_Q$ matrix dimensions, in the $\phi(L_\pi , L_\nu , L)$ basis, for $L=0, 1, 2, 3, 4, 5$ and 6 are 17, 16, 49, 47, 79, 76 and 107 respectively. 
Note that $\lan {\cal C}_2(SU^{(\baa_\pi , \baa_\nu)}(3)) \ran^{(\la , \mu)}$ for $(\la , \mu)=(80,0)$, $(78,1)$, $(76,2)$, $(74,3)$ and $(72,4)$ are 6640, 6400, 6166, 5938 and 5716 respectively. These will allow us to identify the $(\la \mu)K$
for each of the $L$ eigenvector obtained by diagonalization of the $H_Q$ matrices with $\kappa_{\pi \nu}=+1$; see Fig. 2. 

Most important levels in Fig. 2 are the $1^+_S$ and $2^+_S$ levels in the scissors $1^+_S$ band generated by $(78,1)$ irrep. The $M1$ decay of the $1^+_S$ to the ground state $0^+$ is of great interest 
\cite{Scholten,Piet,Kota-book,Hyde}. Also, recently the $E2$ decay of the $1^+_S$ to the lowest $2^+_1$ state in $^{156}$Gd has been measured and in addition, observed is the $2^+_S$ state in this nucleus \cite{Richter}. Following these, studied here are $M1$ and $E2$ strengths from $1^+_S$ and $2^+_S$ levels to the $0^+_g$ and $2^+_g$ levels and also the $E2$ strengths for $4^+_g$ to $2^+_g$ and $2^+_g$ to $0^+_g$ decay. 
     
Using the formulation in Section 4.2, $B(E2)$ and $B(M1)$ values are calculated and the results are given in Table 4 for the four $SU^{(\baa_\pi : \baa_\nu)}(3)$ algebras with $(\baa_\pi) = (\baa_\nu)$ and $\kappa_{\pi \nu}=+1$. Now on, instead of using $(\baa_\pi , \baa_\nu)$ often we will use $(\baa_\pi : \baa_\nu)$.  For the $SU^{(+,+:+,+)}(3)$ situation, as the quadrupole transition operator has same structure as the quadrupole operator in the Hamiltonian in Eq. (\ref{eq.new-15}), analytical formulas can be derived. For the $2^+$ to $0^+$ transition the formula (except for the constant factor and the factor involving effective charges) for the $B(E2)$'s is $4\eta(\eta{N}+3)/(5N)$ and this gives 13.28 for $\eta=4$ and $N=20$. This number is very close to 13.32 in the table. Similarly, for $4^+$ to $2^+$ transition the formula is $8(\eta{N}-2)(\eta{N}+5)/(7 N^2)$ and this gives 18.94 compared to 18.96 in the table. Although all the four algebras give the same spectrum (see Fig. 2), the 
$SU^{(-,+ : -,+)}(3)$ gives weak $E2$ strengths compared to the others. Thus, just as in $sdg$IBM-1 \cite{KSP-19}, again it is possible to have a situation where we have rotational spectrum with weak $E2$ strengths in $sdg$IBM-2. It is also important to mention that the factor '$f$' in Table 4 contains the $N_\pi$ and $N_\nu$ dependence and the remaining factor given in the table is only a function of $N$. For $(\baa_\pi) \neq (\baa_\nu)$, the situation is more complex with the absence of a simple factor $f$ shown in Table 4. For example,
$B(E2; 2^+_g \rightarrow 0^+_g)$ for $SU^{(+,+ : -,-)}(3)$ is $(29.2\, q_\pi - 17.3\, q_\nu)^2$ and similarly, for $1^+_S \rightarrow 2^+_g$ transition it is $(7.89\,q_\pi +
2.89\,q_\nu)^2$.

Turning to $B(M1)$'s, first let us consider the situation with $g^B_\pi =0$ and $g^B_\nu =0$. Then, for the $B(M1)$ for $1^+_S$ to $0^+_g$, formula in the large $N$ limit \cite{Kota-book}, apart from the constant factor $(g_\pi -g_\nu)^2 (3/4\pi)\,\mu_N^2$, is $\sim (2\eta N_\pi N_\nu)/(3N)$. This gives $\sim 13$ compared to $(3.6)^2=12.96$ from the table. More importantly, as seen from the table, the $B(M1)$'s are same for all $(\baa_\pi , \baa_\nu)$. This is due to the fact that the $L_\pi$ (and $L_\nu$) matrix elements do not depend on the $SU(3)$ irreps. Dependence of $B(M1)$'s on $\baa$ arises when $g^B_\pi$ and $g^B_\nu$ are 
non-zero. With these terms present, as can be read off from Table 4, for example 
$$
\barr{l}
B(M1;1^+_S \rightarrow 0^+_g) = \dis\f{3}{4\pi}\;\l(g_\pi -g_\nu\r)^2 \;\l[ 3.6 - 9.52\,g^R_\pi  + 9.38\,g^R_\nu \r]^2\;\mu_N^2 \;; \\
g^R_\pi = \dis\f{8\, g^B_\pi}{(g_\nu -g_\pi)}\;,\;\;\;g^R_\nu = \dis\f{12\, g^B_\nu }{(g_\nu -g_\pi)}
\earr
$$
for $(\baa_\pi : \baa_\nu)=(+,+:+,+)$. In the same way we have from the table,
$$
B(M1;1^+_S \rightarrow 2^+_g) = \dis\f{3}{4\pi}\;(g_\pi -g_\nu)^2 \;\l[2.59 - 3.35\,g^R_\pi + 3.08\,g^R_\nu \r]^2\;\mu_N^2 
$$
for $(\baa_\pi : \baa_\nu)=(+,-:+,-)$ and so on. As seen from the table, depending on the sign of $g^R_\pi$ and $g^R_\nu$, the contribution coming from the $Q \times L$ terms will be substantial or minimal. For example, for $g^R_\pi=g^R_\nu$ the contribution from the $Q \times L$ terms will be nearly zero for $(\baa_\pi : \baa_\nu)=(+,+:+,+)$ while its contribution will be significant for $(\baa_\pi : \baa_\nu)=(-,-:-,-)$. Similarly, if $g^R_\pi > 0$ and $g^R_\nu <0$, then $B(M1)$'s will be enhanced considerably for $(\baa_\pi : \baa_\nu)=(-,-:-,-)$ and reduced for $(\baa_\pi : \baa_\nu)=(+,+:+,+)$. Thus, determination of $g^R_\rho$ from data is important.  

\subsection{$B(E2)$ and $B(M1)$ results from $sdg$IBM-2: $\kappa_{\pi \nu}=-1$}

Considering $\kappa_{\pi \nu}=-1$, for the $N_\pi =8$ and $N_\nu =12$ system, the $(\la , \mu)$ are $(48,32)$, $(47,31)$, $(46,30)$ and so on; see Eq.(\ref{eq.new-16b}). These will give for example the $K$ bands,
$$
(48,32)_{K=0,2,4,\dots}\;,\;\;\;(47,31)_{K=1,3,5,\ldots}\;,\;\;\;(46,30)_{K=0,2,4,\ldots}\;.
$$
The $\lan {\cal C}_2(SU^{(\baa_\pi , \baa_\nu)}(3)) \ran^{(\la , \mu)}$ for the three irreps are 5104, 4861 and 4624 respectively. Energy spectrum generated by $H^{(\baa_\pi , \baa_\nu)}_Q$ is shown in Fig. 3. Only the lowest three bands from the irreps $(48,32)$ and $(47,31)$ are shown in the figure. It is important to note that diagonalization of $H^{(\baa_\pi , \baa_\nu)}_Q$ with $\kappa_{\pi \nu}=-1$ in the $\phi(L_\pi , L_\nu , L)$ basis using the formulation in Section 4.1 generates correctly the eigenvectors 
$$
\l|(48,0)_\nu\; (0,32)_\pi\; (\la,\mu)KL\ran
$$
These are identified using for example the eigenvalues $-5104+(3/4)L(L+1)$, $-4861+(3/4)L(L+1)$ and $-4624+(3/4)L(L+1)$ for the irreps $(48,32)$, $(47,31)$ and $(46,30)$. These are explicitly verified in our calculations.

Spectrum with $\kappa_{\pi \nu}=-1$ is typical of a triaxial rotor. One can also infer this from the value of the shape parameter $\gamma$ given by the $(\la , \mu) = (48,32)$ irrep; see Ref. \cite{JPD-1,Kota-book} for the formula for $\gamma$ in terms of $(\la , \mu)$. It is important to emphasize that the spectrum shown in Fig. 3 is same for the sixteen $SU^{(\baa_\pi , \baa_\nu)}$ algebras with $\kappa_{\pi \nu}=-1$. In $sd$IBM-2, triaxial structure was studied in \cite{DiBi}. Also see \cite{Se-76} for recent efforts in identifying triaxial structures in medium-mass nuclei. 

Going beyond the spectrum,
we have also calculated as an example the $B(M1)$ value for the lowest $1^+_S$ level [this will be uniquely from the $(47,31)$ irrep] to the $0^+_g$ level [this will be uniquely from the irrep
$(48,32)$] using the $T^{M1}$ operator given by Eq. (\ref{eq.new-22}) and the formulation given in Section 4.2 with $g^B_\pi = g^B_\nu=0$. This gives [apart from the $(3/4\pi)\, \mu_N^2$ factor], $6.38\,(g_\pi -g_\nu)^2$ while the simple extension of the formula in \cite{Scholten} gives $6.4\,(g_\pi - g_\nu)^2$. This result is independent of $\baa_\pi$ ($=\baa_\nu$). However, with $g^B_\pi \ne 0$ and $g^B_\nu \neq 0$ formula for the B(M1) is
\be
\barr{l}
B(M1;1^+_S \rightarrow 0^+_g) = \dis\f{3}{4\pi}\;\l(g_\pi -g_\nu\r)^2 \;\l[a + b\,g^R_\pi  + c\,g^R_\nu \r]^2\;\mu_N^2\;;\\
a=2.53,\;\;\;b=-6.69,\;\;\;c=6.59 
\earr \label{eq.new-26}
\ee
for $(\baa_\pi : \baa_\nu)=(+,+:+,+)$. Similarly, for $(+,-:+,-)$, $(-,+:-,+)$ and $(-,-:-,-)$ the values of $(a,b,c)$ in Eq.
(\ref{eq.new-26}) are $(2.53, -3.21, 2.98)$, $(2.53, -1.53, 1.31)$ and $(2.53, 1.95, -2.29)$ respectively.
Due to the degeneracy of levels with $L=2$ and higher (see Fig. 3), further explorations of multiple $SU(3)$ algebras with $\kappa=-1$ need $SU(3) \supset SO(3)$ integrity operators and also the $SU(3)$ irreps here give large number of $K$ values (see Fig. 3). These degeneracies will be lifted and the band structures maintained by adding to $H_Q$ the $SU(3) \supset SO(3)$ integrity basis operators \cite{Kota-book,JPD-1,JPD-2,JPD-3}.

\subsection{$B(E2)$ and $B(M1)$ results from $sdgi$IBM-2}

In order to confirm the generalities of the results obtained for multiple $SU(3)$ algebras using $sdg$IBM-2, calculations are also performed using $sdgi$IBM-2 and here $\eta=6$. As in the previous subsections, again we will restrict to the situation with $\baa_\pi =\baa_\nu$ and this gives eight $SU^{(\baa_\pi , \baa_\nu)}(3)$ algebras for each $\kappa_{\pi \nu}$ value. Note that in the general situation with $\baa_\pi \neq \baa_\nu$, we have a total of 64 $SU(3)$ algebras for each $\kappa_{\pi \nu}$ value. In our numerical study,
considered is a system of six proton bosons and eight neutron bosons so that $N_\pi=6$, $N_\nu=8$ and $N=14$. Now, for $\kappa_{\pi \nu}=+1$, the basis states are $\l|(36,0),L_\pi;(48,0)L_\nu;L\ran$. The $H_Q$ matrix dimensions in this basis for $L=0$, 1, 2 ,3, 4, 5 and 6 are 19, 18, 55, 53, 89, 86 and 121 respectively. Diagonalization of the Hamiltonian given by Eq. (\ref{eq.new-15}), with $\baa_\pi =\baa_\nu$,  gives the eigenstates $\l|(36,0)_\pi (48,0)_\nu (\la,\mu)KL\ran$ for each $\baa_\pi$ choice. The four lowest $(\la , \mu)K$ are
$$
(84,0)K=0,\;\;\;(82,1)K=1,\;\;\;(80,2)K=0,2\;.
$$
Note that $\lan {\cal C}_2(SU^{(\baa_\pi , \baa_\nu)}(3)) \ran^{(\la , \mu)}$ for $(\la , \mu)=(84,0)$, $(82,1)$ and $(80,2)$ are 7308, 7056 and 6810. Using these and the method described in Section 4.1, $(\la ,\mu)$   
associated with the lowest 4 eigenstate for each $L$ are identified. Energy spectrum in $sdgi$IBM-2 is same as the one given in Fig. 2 except that the excitation energies of the $1^+_S$ state and the second $0^+$ state are different. Using the formulation given in Section 4.2, $B(M1)$'s and $B(E2)$'s for the transitions shown in Table 4 are calculated in $sdgi$IBM-2 and the results are shown in Table-5. 
Firstly, we see that the $B(E2)$'s for $SU^{(+,+,+:+,+,+)}(3)$ and $SU^{(+,+,-: +,+,-)}(3)$ are much stronger than those from the other $SU^{(\baa_\pi , \baa_\nu)}(3)$. Also, it is easy to see that the results for $B(E2)$ and $B(M1)$ values for $SU^{(+,+,+ : +,+,+)}(3)$ agree with the analytical formulas given in Section 4.3. It is useful to note that the $B(E2)$'s are smallest for $SU^{(+,-,- : +,-,-)}(3)$. In addition, as in $sdg$IBM-2, the $B(M1)$'s will be small or large depending on the signs of $g^R_\pi$ and $g^R_\nu$. Thus, the $B(E2)$ and $B(M1)$ structure of the levels of the scissors $1^+$ band depends on $\baa$ and it is possible to have the $1^+_S$ band with the $E2$ and $M1$ decay of the lowlying levels of this band to the $g$ band strong or weak. Experimental data such as those reported in \cite{Richter} if accumulated for more deformed nuclei, will be clearly useful in deciding which $\baa$ is appropriate for a given nucleus. It is important to note that the $sdgi$IBM-2 results in Table 5 confirm generality of the results obtained using $sdg$IBM-2.

Turning briefly to the situation with $\kappa_{\pi \nu}=-1$, for the $N_\pi =6$ and $N_\nu =8$ system, the $(\la , \mu)$ are $(48,36)$, $(47,35)$, $(46,34)$ and so on. Diagonalization of $H^{(\baa_\pi , \baa_\nu)}_Q$ with $\kappa_{\pi \nu}=-1$ in the $\phi(L_\pi , L_\nu , L)$ basis generates the eigenvectors 
$$
\l|(48,0)_\nu\; (0,36)_\pi\; (\la,\mu)KL\ran
$$
These are identified using for example the eigenvalues $-5580+(3/4)L(L+1)$, $-5325+(3/4)L(L+1)$ and $-5076+(3/4)L(L+1)$ for the irreps $(48,36)$, $(47,35)$ and $(46,34)$. These are explicitly verified in our calculations. Just as before, the spectrum is same for all $SU^{(\baa_\pi ,\baa_\nu)}(3)$ algebras. 
Note that the lowest $1^+_S$ level here is generated by the irrep $(47,35)$  and the $0^+_g$ level by the irrep $(48,36)$. The 
$B(M1)$ values for the decay of the $1^+_S$ level to the $0^+_g$ level are calculated as before and the result for $(\baa_\pi : \baa_\nu)=(+,+,+:+,+,+)$ is
\be
\barr{l}
B(M1;1^+_S \rightarrow 0^+_g) = \dis\f{3}{4\pi}\;\l(g_\pi -g_\nu\r)^2 \;\l[a + b\,g^R_\pi  + c\,g^R_\nu \r]^2\;\mu_N^2\;;\\
a=2.61,\;\;\;b=-10.34,\;\;\;c=10.23 
\earr \label{eq.new-27}
\ee
Similarly, for $(+,+,-:+,+,-)$, $(+,-,+:+,-,+)$, $(+,-,-:+,-,-)$, $(-,+,+:-,+,+)$, $(-,+,-:-,+,-)$, $(-,-,+:-,-,+)$ and $(-,-,-:-,-,-)$ the values of $(a,b,c)$ in Eq.
(\ref{eq.new-27}) are $(2.61, -8.83, 8.64)$, $(2.61, -4.23, 3.93)$, $(2.61, -2.72, 2.34)$, $(2.61, -4.55, 4.32)$, $(2.61, -3.05, 2.73)$, $(2.61, 1.56, -1.98)$ and $(2.61, 3.06, -3.57)$ respectively. Because of the degeneracies in the
spectrum, other $B(M1)$ and $B(E2)$ values are not listed here. As stated before, for complete understanding of multiple $SU(3)$ algebras with $\kappa_{\pi \nu}=-1$, we need integrity basis spectroscopy. 

In concluding the analysis in IBM spaces, it is important to underline the key role played in Sections 2 and 4 by the analytical results, valid for sufficiently large $N$ (boson number) values, derived in \cite{KM-1}. 
  
\begin{table}
\caption{$B(E2; L_i \rightarrow L_f)$ and $B(M1; L_i \rightarrow L_f)$  values for some transitions involving $g$ ($0^+$) band and $K=1^+_S$ band levels for a $N_\pi=6$ and $N_\nu=8$ system generated by the $SU^{(\baa_\pi , \baa_\nu)}(3)$ algebras with $\baa_\pi = \baa_\nu$ in $sdgi$IBM-2. The $B(E2; L_i \rightarrow L_f)$ values in the table are in units of $(5/16\pi)\,b^4$. $B(M1; L_i \rightarrow L_f)$ is $\l[R(M1; L_i \rightarrow L_f)\r]^2\; (3/4\pi)\,\mu_N^2$. Note that each $R(M1)$ is a sum of three terms given in the table. The numbers in columns 3-10 are for the eight values of $\baa_\pi$ and they have to be multiplied by the factor $f$ given in the last column. In the table $1^+_S$ and $2^+_S$ are the $1^+$ and $2^+$ levels of the $1^+_S$ band. See text for more details.}

\label{table4}
{\scriptsize{
\begin{tabular}{ccccccccccc}
\hline
$transition$ & $\;\;\;\;\;$ & $(+,+,+)$ & $(+,+,-)$ & 
$(+,-,+)$ & $(+,-,-)$ & $(-,+,+)$ & $(-,+,-)$ & $(-,-,+)$ & $(-,-,-)$ & $f$  \\
\hline
$2^+_g \rightarrow 0^+_g$ & $B(E2)$ & $ 29.85$  & $ 20.68$  & $  3.69$  & $  1.01$  & $  4.75$  & $  1.60$  & $  1.86$  & $  5.19$  & $\l(6\,q_\pi + 8\, q_\nu \r)^2$ \\  
$4^+_g \rightarrow 2^+_g$ & $B(E2)$ & $ 42.58$  & $ 29.39$  & $  5.24$  & $  1.40$  & $  6.83$  & $  2.28$  & $  2.63$  & $  7.44$  & $\l(6\, q_\pi + 8\, q_\nu\r)^2$ \\
$1^+_S \rightarrow 2^+_g$ & $B(E2)$ & $ 64.86$  & $ 48.53$  & $  9.15$  & $  3.75$  & $  9.22$  & $  3.80$  & $  3.97$  & $  9.48$  & $\l(q_\pi - q_\nu\r)^2$ \\ 
$2^+_S \rightarrow 2^+_g$ & $B(E2)$ & $  9.04$  & $  6.23$  & $  1.77$  & $  0.68$  & $  2.51$  & $  1.15$  & $  0.01$  & $  0.36$   & $\l(q_\pi - q_\nu\r)^2$ \\
$2^+_S \rightarrow 0^+_g$ & $B(E2)$ & $ 24.46$  & $ 15.61$  & $  2.61$  & $  0.39$  & $  4.35$  & $  1.19$  & $  1.55$  & $  5.01$   & $\l(q_\pi - q_\nu\r)^2$ \\
\hline
\\
$1^+_S \rightarrow 0^+_g$ & $R(M1)$ & $3.72$  & $ 3.72$  & $3.72$  & $3.72$  & $3.72$  & $3.72$  & $3.72$  & $3.72$  & $(g_\nu -g_\pi)$ \\
& & $-14.71$  & $-12.55$  & $-6.27$  & $-4.11$  & $-6.85$  & $  -4.69$  & $ 1.59$  & $ 3.75$  & $6\, g^B_\pi$ \\
& & $14.56$  & $12.29$  & $5.73$  & $3.46$  & $6.36$  & $4.09$  & $-2.47$  & $-4.74$  & $8\, g^B_\nu$ \\  
$1^+_S \rightarrow 2^+_g$ & $R(M1)$ & $ 2.68$  & $ 2.68$  & $ 2.68$  & $ 2.68$  & $ 2.68$  & $ 2.68$  & $ 2.68$  & $ 2.68$  &  $(g_\nu -g_\pi)$ \\
& & $-10.58$  & $-9.03$  & $-4.51$  & $-2.96$  & $-4.93$  &
$-3.37$  & $ 1.15$  & $ 2.70$  & $6\,g^B_\pi$ \\
& & $10.48$  & $8.84$  & $4.12$  & $2.49$  & $4.56$  & $2.94$  & $-1.78$  & $-3.41$  & $8\, g^B_\nu$ \\  
$2^+_S \rightarrow 2^+_g$ & $R(M1)$ &  $ 4.58$  & $ 4.58$  & $ 4.58$  & $ 4.58$  & $ 4.58$  & $ 4.58$  & $ 4.58$  & $ 4.58$  & $(g_\nu -g_\pi)$ \\
& & $-18.12$  & $-15.45$  & $-7.76$  & $-5.10$  & $-8.5$  & $  -5.84$  & $ 1.85$  & $ 4.51$  & $6\, g^B_\pi$ \\
& & $17.93$  & $15.13$  & $7.1$  & $4.3$  & $7.9$  & $5.1$  & $  -2.93$  & $-5.73$  & $8\, g^B_\nu$ \\  
\hline
\end{tabular}
}}
\end{table}

\begin{figure}
\includegraphics[width=0.6\linewidth]{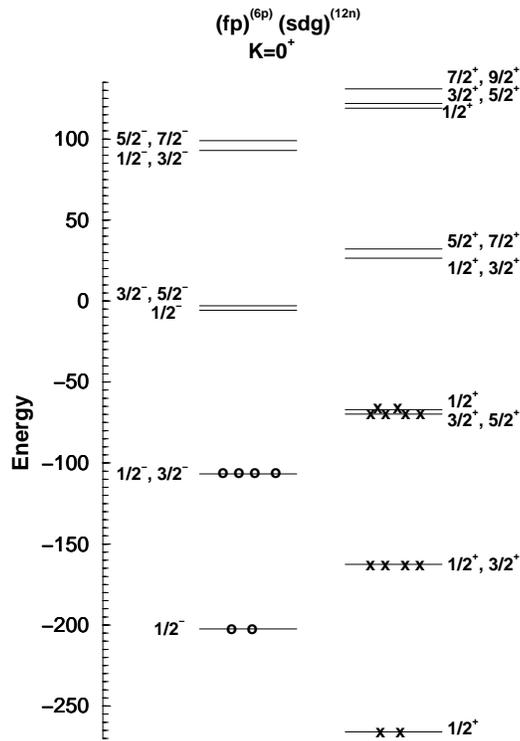}
\caption{Hartree-Fock sp spectrum and the lowest intrinsic state for the 
$(fp)^{6_p} (sdg)^{12_n}$ system generated by the eight $H_Q$ operators in Eq.
(\ref{eq.abcd}).  In the figure, the symbol $\times$ denotes neutrons and
$\mbox{o}$
denotes protons. Shown in the figure are the $k$ values of the sp orbits and
each orbit is doubly degenerate with $\l|k\ran$ and $\l|-k\ran$ states. The
spectra  are same for all  the eight Hamiltonians although the sp wavefunctions
are different. Note that the energies in the figures are unitless and the  unit
MeV has to be put back after multiplying with an appropriate scale factor  if
the results are used for a real nucleus.} 
\label{fig-3}  
\end{figure}
     
\section{Multiple $SU(3)$ algebras in shell model: DSM results for scissors $1^+$ band from $(fp)^{6p}(sdg)^{12n}$ system}

In general in shell model for a nucleus, with valence protons ($p$) say in a oscillator shell $\eta$ and valence neutrons ($n$) in a $\eta^\pr$ shell, the quadrupole operator is given by extending Eq. (\ref{eq.new-13}) and applying Eq. (\ref{eq.new-1}) in proton and neutron shell model orbital spaces. Then, 
\be
Q^2(\baa_p : \baa_n ; \kappa_{pn}) = Q^2_p(\baa_p) + \kappa_{pn } Q^2_n(\baa_n)\;;\;\;\;\kappa_{pn}=\pm 1\;.
\label{eq.new-28}
\ee
With this, the number of $SU(3)$ algebras is $2^{\l[\f{\eta}{2}\r]} \times 2^{\l[\f{\eta^\pr}{2}\r]}$ for each $\kappa_{\pi \nu}$ value. In this Section, we will present some results for multiple $SU(3)$ algebras in SM using the example of a system of 6 protons in $pf$-shell ($\eta=3$) and 12 neutrons in $sdg$-shell ($\eta^\pr =4$). This gives $2 \times 4 =8$ $SU^{(\baa_p : \baa_n)}(3)$ algebras or eight $Q.Q$ Hamiltonians, 
\be
H^{(\baa_p : \baa_n)}_Q = -\dis\frac{1}{4}\; Q^2(\baa_p : \baa_n ; \kappa_{pn}) \cdot Q^2(\baa_p : \baa_n ; \kappa_{pn})
\label{eq.abcd}
\ee
for each $\kappa_{pn}$ value, all giving the same spectrum but different properties for $B(E2)$'s and $B(M1)$'s. The SM matrix dimensions for $(fp)^{m_p=6,S_p=0} (sdg)^{m_n=12,S_n=0}$ system are very large and therefore full SM calculations are not yet possible. Instead, here we employ DSM by considering some limited number of configurations using deformed sp states (see \cite{KS-book} for details regarding DSM). Before discussing the results from DSM, it is useful to mention that in
the $(fp)^{m_p=6,S_p=0} (sdg)^{m_n=12,S_n=0}$ example, the lowest $K_p=0$ and $K_n=0$ bands belong to the $SU(3)$ irreps $(12,0)$ and $(24,0)$ respectively for $\kappa_{pn}=+1$. Note that with more general proton and neutron numbers, the lowest $SU(3)$ irreps (giving $K=0^+$ bands) will not be $(\la ,0)$ type (see for example Section 3 and Tables 3.1-3.3 in \cite{Kota-book}). Now, the $SU^{(\baa_p : \baa_n)}(3)$ irreps are
\be
(\la , \mu)_K = (36,0)_{K=0} \oplus (34,1)_{K=1} \oplus (32,2)_{K=0,2} \oplus \ldots
\label{eq.new-29}
\ee
The $\lan {\cal C}_2(SU^{(\baa_\pi : \baa_\nu)}(3)) \ran^{(\la , \mu)}$ for $(\la , \mu)=(36,0)$, $(34,1)$ and $(32,2)$ are 1404,
1296 and 1194 respectively. These values are used to identify the bands given by Eq. (\ref{eq.new-29}).

\begin{table}
\caption{$B(E2; L_i \rightarrow L_f)$ and $B(M1; L_i \rightarrow L_f)$ values for some transitions involving $g$ ($0^+$) band and $K=1^+_S$ band levels for a shell model$(fp)^{m_p=6} (sdg)^{m_n=12}$  system generated by the eight $SU^{(\baa_p : \baa_n)}(3)$ algebras. First row in the table gives the eight $(\baa_p : \baa_n)$; note that $\baa_p=\pm$ and $\baa_n = (\pm , \pm )$. The $B(E2)$ values are in units of $e^2fm^4$ and they are calculated using $b=2.07\;fm$ and effective charges $e_p=1.5e$ and $e_n=0.5e$. Similarly, $B(M1; L_i \rightarrow L_f)$ = $\l[R(M1; L_i \rightarrow L_f)\r]^2\; (3/4\pi)\,\mu_N^2$ and $R(M1)$ is given in the table. Note that $R(M1)$ is a sum of the three terms given in the table and they have to be multiplied by the factor $f$ given in the last column. In the table $1^+_S$ and $2^+_S$ are the $1^+$ and $2^+$ levels of the $1^+_S$ band. See text for more details.}
{\scriptsize{
\begin{tabular}{ccccccccccc}
\hline
$transition$ & $\;\;\;\;\;$ & $(-:-,-)$ & $(-:-,+)$ & $(-:+,-)$ & $(-:+,+)$ & $(+:-,-)$ & $(+:-,+)$ & $(+:+,-)$ & $(+:+,+)$ & $f$ \\
\hline
$2_g^+\rightarrow 0^+_g$ & $B(E2)$ 
	     &1415   &603   &1046 &372  &124  &3.6   &34   &52 &    \\
$4_g^+\rightarrow 2^+_g$ & $B(E2)$ 
	     &2007   &861   &1481 &530  &180  &4.1   &51   &70 &   \\
$1_S^+\rightarrow 2^+_g$ & $B(E2)$ 
	     &20.4   &62    &33   &1.4  &52   &14.7  &34   &6.6 &  \\
$2_S^+\rightarrow 0^+_g$ & $B(E2)$ 
	     &36     &20    &13   &29.4 &13   &9.5   &7.4  &0.8 &   \\
$2_S^+\rightarrow 2^+_g$ & $B(E2)$ 
	     &1      &6     &2    &8.0  &6    &0.9   &4.0  &0.3 & \\
\hline
$1_S^+\rightarrow 0^+_g$ & $R(M1)$ & 0.68 & 0.68 & 0.68 & 0.68 & 0.68 & 0.68 & 0.68 & 0.68 & 1 \\
& & $-7.51$ & 7.51 & $-7.51$ & 7.51 & 1.3 & $-1.3$ & 1.3 & $-1.3$ & $g_p^c$ \\
& & 15.02 & $-2$ & 9.76 & 3.27 & 15.02 & $-2$ & 9.76 & 3.27 & $g_n^c$ \\
$1_S^+\rightarrow 2^+_g$ & $R(M1)$ & 0.5 & 0.5 & 0.5 & 0.5 & 0.5 & 0.5 & 0.5 & 0.5 & 1 \\
& & 14.51 & $-3.88$ & $-3.88$ & 14.51 & 6.37 & 6.37 & 4.27 & 4.27& $g_p^c$ \\
& & $-30.21$ & $-7.75$ & $ 1.70$ & $-7.77$ & 8.20 & $-7.75$ & $-23.72$ & $-7.77$ & $g_n^c$ \\
$2_S^+\rightarrow 2^+_g$ & $R(M1)$ & 0.83 & 0.83 & 0.83 & 0.83 & 0.83 & 0.83 & 0.83 & 0.83 & 1 \\ 
& & 8.77 & $-8.76$ & $-8.76$ & 9.07 & 1.42 & 1.42 & $-1.42$ & $-1.42$ & $g_p^c$ \\
& & $-17.98$ & 2.27 & 11.71 & 4 & 17.98 & 2.27 & $-11.71$ & 4 & $g_n^c$ \\                                         
\hline
\end{tabular}
}}
\label{tab-5}
\end{table}

Starting with $H^{(\baa_p : \baa_n)}_Q$ with $\kappa_{pn}=+1$, first the sp energies and the two-body matrix elements of the $pp$ and $nn$ parts and similarly the two-body matrix elements of the $pn$ part of $H_Q$ are obtained for each $(\baa_p :\baa_n)$ (see \cite{EPJST} for the formulation to obtain these). Using these, HF calculations assuming axial symmetry are carried out. The HF sp spectrum and the lowest intrinsic state obtained are shown in Fig. 4. It is important to note that the sp spectrum is same for the eight $H_Q$ operators. However, the sp wavefunctions differ in phases of their various components (see for example Table 3 in \cite{EPJST}). Therefore, the intrinsic quadrupole moments of the lowest intrinsic state generated by the eight $SU(3)$ algebras are different. The quadrupole moments (in units of $b^2$ where $b$ is oscillator length parameter) for $SU^{(\baa_p : \baa_n)}(3)$ with $(\baa_p : \baa_n)=(-:-,-)$, $(-:-,+)$, $(-:
+,-)$, $(-:+,+)$, $(+:-,-)$, $(+:-,+)$, $(+:+,-)$ and $(+:+,+)$ are $71.96$, $31.43$, $55.02$, $14.50$, $44.65$, $-24.68$, $27.72$ and $-12.81$ respectively. Thus, the shapes are prolate for all except for $(+:-,+)$ and $(+:+,+)$. It is easy to see that the lowest intrinsic state gives the $g$ ($K=0^+$) band (see Fig. 4)and its 
mixing with other excited intrinsic states is very small. As we are interested
in the scissors $1^+_S$ band, starting with the lowest intrinsic state with $K=0^+$, we have considered six excited intrinsic states with $K=1^+$ by exciting a proton or a neutron from the highest filled orbits to the next empty orbits (see Fig. 4). These configurations are $(1/2_2)^{-p} (1/2_3)^p$, $(3/2_1)^{-p} (1/2_3)^n$, $(1/2_3)^{-n} (1/2_4)^n$, $(1/2_3)^{-n} (3/2_3)^n$, $(3/2_2)^{-n} (1/2_4)^n$, $(5/2_1)^{-n} (3/2_3)^n$. Performing angular momentum projection and band mixing, the $K=1^+_S$ band is generated. We have verified that the energy of the $0^+$ level of the lowest $K=0^+$ band (i.e. $g$ band) differs from the expected value from the $(36,0)$ irrep [see Eq. (\ref{eq.new-29})] by less than 1\%. Similarly, we have also verified that the $1^+$ level of the $1^+_S$ band differs from the value expected from $(34,1)$ irrep by less than 1\%. 
It is seen that the structure of the $K=1^+$ band that is identified as the $1^+_S$ band is generated mainly by the deformed 
configurations obtained by promoting a proton from occupied $k=1/2^-$ or
$3/2^-$ orbit to unoccupied $k=1/2^-$. The $1^+$level of this band shows 
strong $B(M1)$ transition to the levels of the ground band ($1^+$ level of the other $K=1^+$ bands are found to give very small $B(M1)$ values).
We have also verified that the levels of the $g$ and the $1^+_S$ band follow $L(L+1)$ law very closely. Thus, DSM produces to good accuracy exact $SU(3)$ symmetry results for spectra. Though the spectra from the eight
$SU^{(\baa_p , \baa_n)}(3)$ algebras are same, the $B(E2)$ and $B(M1)$ properties are expected to be different (we have already seen this in the intrinsic quadrupole moments).            

Table 6 gives the calculated results for $B(E2)$'s and $B(M1)$'s involving some of the low-lying levels in $g$ band and $1^+_S$ band. The $E2$ operator used is 
\be
T^{E2}= [e_p Q_p^2(-) + e_n Q_n^2(-,-)]\;b^2
\label{eq.sm-q2}
\ee
where $e_p$ and $e_n$ are proton and neutron effective charges and $b$ is the oscillator length parameter (as stated in Section 3, the above choice is standard in shell model (and DSM) studies \cite{SM-MM1,SM-MM2,KS-book} and also validated by the DSM results in Appendix-A). Similarly, the $M1$ operator used, following Eq. (\ref{eq.new-22}), is
\be
\barr{l}
T^{M1} = T^{M1}_{sp} + T^{M1}_{coll}\;;\\
T^{M1}_{sp} = [g^p_\ell \vec{L_p} + g^n_\ell \vec{L_n} + g^p_s \vec{S_p} + g^n_s \vec{S_n}]\;\mu_N\;,\\
T^{M1}_{coll} = \l[g^c_p\;\l(Q^2_p(-) \times L^1_p\r)^1_q + 
g^c_n\;\l(Q^2_n(-,-) \times L^1_n\r)^1_q\r]\;\mu_N\;.
\earr \label{eq.sm-m1}
\ee
Note that $T^{M1}_{sp}$ is the standard one-body $M1$ operator with bare $g$-factors, i.e. $g^p_\ell=1$, $g^n_\ell=0$, $g^p_s=5.586$ and $g^n_s=-3.826$. However, this gives results independent of the phases $(\baa_p : \baa_n)$ as can be seen from Table 6 just as in $sdg$IBM-2 and $sdgi$IBM-2. In order to get phase dependence, we have added as in Eq. (\ref{eq.new-22}) the collective part $T^{M1}_{coll}$. With Eq. (\ref{eq.sm-m1}) the $B(M1)$'s are given by
\be
\barr{rcl}
B(M1; J_i \rightarrow J_f) & = & \dis\f{3}{4\pi}\;
\l|R(M1; J_i \rightarrow J_f)\r|^2\;\mu_N^2\;; \\
R(M1; J_i \rightarrow J_f) & = & X_{sp} + g^c_p X_p + g^c_n X_n
\earr \label{eq.sm-m1a}
\ee
where, with $\hat{J}_i=\sqrt{2J_i+1}$,
\be
\barr{rcl}
X_{sp} & = & (\hat{J}_i)^{-1}\;\lan J_f \mid\mid T^{M1}_{sp} \mid\mid J_i\ran \\
X_p & = & (\hat{J}_i)^{-1}\;\lan J_f \mid \mid \l(Q^2_p(-) \times L^1_p\r)^1 \mid\mid J_i\ran \;,\\ 
X_n & = & (\hat{J}_i)^{-1}\; \lan J_f \mid\mid \l(Q^2_n(-,-) \times L^1_n\r)^1 \mid\mid J_i\ran \;.
\earr \label{eq.sm-m1b}
\ee
It is in general easy to calculate $X_{sp}$ and for $X_p$ (and similarly for $X_n$) we use the formula,
\be
\barr{l}
X_p = \dis\f{\dis\sqrt{3}\,(-1)^{J_i+J_f+1}}{\dis\sqrt{(2J_i+1)}} \dis\sum_{J^\prime}\;\l\{\barr{ccc} 2 & 1 & 1 \\ J_i & J_f & J^\prime \earr \r\} \\
\times\; \lan J_f \mid\mid Q^2_p(-) \mid\mid J^\prime\ran\;
\lan J^\prime \mid\mid L_p^1 \mid\mid J_i\ran\;.
\earr \label{eq.sm-m1c}
\ee
In Table 6, given are the values of $X_{sp}$, $X_p$ and $X_n$ for three transitions and using them it is easy to read off the $B(M1)$ values. For example for $(\baa_p : \baa_n)=(-:-,-)$ we have
$$
B(M1; 1^+_S \rightarrow 0^+_g)= \dis\f{3}{4\pi} \l[0.66 - 7.51\,g^c_p  + 15.02\,g^c_n \r]^2\;\mu_N^2
$$
and similarly, for $(\baa_p : \baa_n)=(+:-,-)$ 
$$
B(M1; 1^+_S \rightarrow 2^+_g)= \dis\f{3}{4\pi} \l[0.5 + 6.37\,g^c_p  + 8.2\,g^c_n \r]^2\;\mu_N^2\;.
$$

It is seen from Table 6 that the trends in $B(E2)$'s is similar to the trends seen in IBM-2 results shown in Tables 4 and 5.
The $B(E2)$'s depend strongly on $(\baa_p : \baa_n)$ and for $(-:-,-)$ they follow from $SU(3)$ algebra as the $E2$ operator for this situation is a generator of $SU^{(\baa_p : \baa_n)}(3)$ while it is not a generator of the remaining seven $SU(3)$ algebras. Turning to $B(M1)$, in the situation $g^c_p =g^c_n=0$ as the parts with $Q^2$ are absent in the $M1$ operator, just as in IBM-2 the $B(M1)$'s are same for all the eight choices of
$(\baa_p : \baa_n)$. The full $M1$ operator with $T^{M1}_{coll}$  
generates $(\baa_p : \baa_n)$ dependence in $B(M1)$'s. Depending on the signs of $g^c_p$ and $g^c_n$, the $B(M1)$'s will be large or small and it is easy to see that there is strong dependence on
$(\baa_p : \baa_n)$. This may help in identifying the presence of
multiple $SU(3)$ algebras in nuclear data.
 
Although we have carried out angular momentum projection and band mixing using the intrinsic states of the total $pn$ system,
there is an alternative approach to generate the $1^+_S$ band as described in \cite{BPM,Sun} and this is similar to the approach used in Section 4. In summary, although the SM results (obtained using DSM) are consistent with the results from IBM-2 reported in Section 4, clearly more extensive SM studies are needed for arriving at more conclusive results for the $E2$ and $M1$ properties of the $1^+_S$ band generated by multiple $SU(3)$ algebras in SM spaces. 
 
\section{conclusions and future outlook}

Going beyond the first studies reported in \cite{KSP-19,EPJST} by Kota {\it et al.} for multiple $SU(3)$ algebras in IBM and SM spaces, where only the $g$ ($K=0^+$) band in various rotational systems are analyzed, in the present paper results are presented for the $E2$ decay properties of the levels of $\beta$ and $\gamma$ bands and $E2$ and $M1$ decay properties of the levels of the $1^+_S$ scissors band in heavy deformed nuclei as generated by multiple $SU(3)$ algebras. Results from IBM are presented in Sections 2 and 4 and those from SM are presented in Sections 3 and 5.

Properties of $\gamma$ and $\beta$ band levels generated by multiple $SU^{(\baa)}(3)$ algebras in $sdg$IBM and $sdgi$IBM are studied first by deriving the structure coefficients of the intrinsic states that generate the $g$, $\beta$ and $\gamma$ bands by solving an eigenvalue equation. Importantly, the structure coefficients are determined as a function of $(\baa)$ as given by Eqs. (6) and (7). It is important to note that in IBM the $g$ band is generated by $(\eta N,0)$ irrep and the $(\beta , \gamma)$ bands by the irrep $(\eta N-4,2)$. In terms of the structure coefficients and with angular momentum projection, formulas (though complex) for the matrix elements of
the quadrupole operator (both for inter and intra band transitions) are derived many years back in \cite{KM-1}. Applying these [see Eqs. (10,11,20)] $g \rightarrow g$,
$\beta \rightarrow g$ and $\gamma \rightarrow g$ $E2$ transition strengths are studied and the results are presented in Tables 1 and 2. These clearly establish that the $E2$ decay of the levels of $\beta$ and $\gamma$ bands to the ground band are quite different for some of the $SU^{(\baa)}(3)$ algebras with strong dependence on ($\baa$). Going further, a shell model example is also considered for the study of $\gamma$ band properties generated by multiple $SU(3)$ algebras. In SM the situation is more complex as often both the $g$ ($K=0$) and $\gamma$ ($K=2$)
band are generated by the same $SU(3)$ irrep. As seen from Table 3, SM results are consistent with the conclusions from IBM. However, as pointed out in Section 3 many more studies of the $\gamma$ band (and higher $K$ bands) in SM are needed for more firm conclusions.

Investigating multiple $SU(3)$ algebras further, we have considered in Section 4  $1^+_S$ (scissors) band in heavy deformed nuclei that needs IBM-2, i.e. proton-neutron IBM (note that we have used IBM-1 in Section 2 and also in our previous publications \cite{KSP-19,EPJST}). We have used the fact that it is easy to identify the $g$ ($K=0$) band and the scissors $1^+_S$ band by diagonalizing the $Q.Q$ Hamiltonian ($Q=Q_\pi \pm Q_\nu$) in the product basis generated by the $g(K=0)$ bands in proton ($\pi$) and neutron ($\nu$) boson spaces [see Eqs. (15,19-21)]. In addition used is the formulation in \cite{KM-1} in $\pi$ space with ($\baa_\pi$) and $\nu$ space with ($\baa_\nu$). Let us add that in most of the calculations we have considered $Q=Q_\pi + Q_\nu$. The simple $E2$ transition operator given by Eq. (24) generates $(\baa)$ dependence of $E2$ strengths. However, for $M1$ strengths we found that it is necessary to include $(Q \times L)^1_q$ type terms in the $M1$ operator and these terms generate $(\baa)$ dependence of $M1$ strengths. Results for $E2$ and $M1$ decay properties of the $1^+$ and $2^+$ levels of the $1^+_S$ band are
given in Table 4 for $sdg$IBM-2 and in Table 5 for $sdgi$IBM-2. 
These show that it is possible to have the $1^+_S$ band with the $E2$ and $M1$ decay of the low-lying levels of this band to the $g$ band strong or weak depending on ($\baa$). Going further, a shell model example, with six protons in $(pf)$ shell and twelve neutrons in $(sdg)$ shell, is also considered for the study of $1^+_S$ band properties generated by multiple $SU(3)$ algebras. The SM results presented in Table 6 are consistent with the observation from $sdg$IBM-2 and $sdgi$IBM-2 analysis.

In the analysis presented in the present and previous papers for different realizations of $SU(3)$ in a $\eta$ shell, quadrupole moments and $B(E2)$ vales (in some situations also $B(M1)$'s) generated by the quadrupole-quadrupole Hamiltonians based on all these realizations are examined using a $E2$ transition operator with a fixed $(\baa)$. Easy to understand $sd$IBM example presented in Appendix-A is in conformity with this. For further understanding of this, a $(1f2p)$ shell model example is also presented in the Appendix. Let us add that the significance of multiple $SU(3)$ algebras follow from the results in Tables 1-6. Further deeper understanding may follow by using most general coherent states in $sdg$ and $sdgi$ spaces but such a study is beyond the scope of the present paper.  

All the results presented in this paper establish the following generic features: (i) with multiple $SU(3)$ algebras, it is possible to have rotational bands with some $SU(3)$ algebras giving very weak $E2$ strengths and some others the usual strong strengths; (ii) $E2$ decay strengths of the levels of $\beta$ and $\gamma$ bands to the ground band levels depend strongly on $\baa$ of $SU^{(\baa)}(3)$ algebras and thus for example with $SU(3)$ symmetry it is possible to have situations with strong $\gamma \rightarrow g$ $E2$ transition strengths; (iii) it is possible to have the $1^+_S$ band in heavy deformed nuclei with the $E2$ and $M1$ decay of the low-lying levels of this band to the $g$ band strong or weak depending on $(\baa)$; (iv) all $SU(3)$ algebras generate the same spectrum but different $E2$ and $M1$ decay characteristics. These may help in the examination of experimental data for the presence of multiple $SU(3)$ algebras in nuclei. Finally, we speculate that the prolate-oblate transition (with excitation energy) seen in some nuclei in mass 60-90 region \cite{KS-book,Zuk} such as $^{72}$Kr \cite{Kr72} and some of the so called magnetic rotation bands \cite{Mag-1,Mag-2} and chiral bands \cite{Mag-3} could be understood as a manifestaion of some of the multiple $SU(3)$ algebras discussed in this paper.

\section{Acknowledgments}

R. Sahu is thankful to SERB of Department of
Science and Technology (Government of India) for financial support. Thanks are due to P.C. Srivastava for some useful correspondence.

\appendix

\renewcommand{\theequation}{A\arabic{equation}}
\setcounter{equation}{0}

\section{}

Let us consider $sd$IBM with $s$ ($\ell=0$) and $d$ ($\ell=2$) bosons that is simple to analyze. Here, $\eta=2$ and there will be two $SU(3)$ algebras, $SU^{(+)}(3)$ and 
$SU^{(-)}(3)$ as seen from the discussion in Section 2 with $\alpha_{02}=\pm 1$. 
Correspondingly, we have $H_Q^{(+)}$ a $H_Q^{(-)}$ Hamitonians [see Eq. (2)]. 
Note that from Eq. (1) we have
\be
Q^2_\mu(\alpha_{02}) = 2\dis\sqrt{2} \l[-\dis\f{\dis\sqrt{7}}{2}\,\l(d^\dagger 
\tilde{d}\r)^2_\mu + \alpha_{02}\;\l(s^\dagger \tilde{d} + 
d^\dagger \tilde{s}\r)^2_\mu \r]\;.
\ee
Using this and solving Eq. (5) gives the intrinsic state for the ground $K=0$ 
band generated by $H_Q^{(\alpha_{02})}$,
\be
b^\dagger_{0_g}\l.\l|0\r.\ran = \dis\f{1}{\dis\sqrt{3}}\,s^\dagger_0 
\l.\l|0\r.\ran + \alpha_{02}\;\dis\sqrt{\dis\f{2}{3}}\,d^\dagger_0
\l.\l|0\r.\ran\;;\;\;\;\alpha_{02}=\pm 1\;.
\ee
Now, choosing the $E2$ transition operator to be $T^{E2}=q_2 Q^2_\mu(+)$ and 
calculating the quadrupole moments $Q(2^+_1)$ and $Q(4^+_1)$ of the $2^+_1$ and 
$4^+_1$ states of the ground $K=0$ band and also $B(E2; 2^+_1 \rightarrow 0^+_1)$ 
and $B(E2; 4^+_1 \rightarrow 2^+_1)$ using Eqs. (A2) and (10), the following 
results are obtained. For $H_Q^{(+)}$ we have, 
\be
\barr{l}
Q(2^+_1)= -21,\;\;\;\;\; Q(4^+_1)= -27, \\
B(E2; 2^+_1 \rightarrow 0^+_1) = 110, \;\;\;\;\;
B(E2; 4^+_1 \rightarrow 2^+_1) = 154 
\earr
\ee
with the $Q$'s in $q_2$ units and $B(E2)$'s in $(q_2)^2$ units.
Similarly, for $H_Q^{(-)}$ we have, 
\be
\barr{l}
Q(2^+_1)= 7,\;\;\;\;\; Q(4^+_1)= 8,\\
B(E2; 2^+_1 \rightarrow 0^+_1) = 15,\;\;\;\;\; 
B(E2; 4^+_1 \rightarrow 2^+_1) = 20\;.
\earr
\ee
Thus, as expected, $H_Q^{(+)}$ generates prolate shape and 
$H_Q^{(-)}$ generates oblate shape. On the other hand, if we change the $E2$ 
transition operator to $T^{E2}=q_2 Q^2_\mu(-)$, the $H_Q^{(+)}$ gives $Q(2^+_1)= 
7\,q_2$ and $Q(4^+_1)= 8\;q_2$. Similarly, $H_Q^{(-)}$ gives $Q(2^+_1)= -21\,q_2$, $Q(4^+_1)= -27\,q_2$. As seen from these results, clearly when we use $H_Q(\baa)$ 
with $(\baa)$ changing, the form of the $E2$ transition operator should not change -
otherwise in the present example prolate nuclei become oblate and oblate nuclei become prolate.
In IBM literature \cite{Iac}, for $SU(3)$ limit the choice is correctly $T^{E2}=q_2 
Q^2_\mu(+)$ so that $H_Q^{(+)}$ gives prolate and $H_Q^{(-)}$ oblate shapes.
These results apply also to fermion systems as demonstrated with a $(1f2p)$ 
shell model example below.

Let us consider six protons in $(1f2p)$ shell (then $\eta=3$). The ground $K=0$ band here is generated by the $SU(3)$ irrep$(12,0)$. With $\eta=3$, we have two $SU(3)$ algebras with $\alpha_{13}=\pm 1$ giving $SU^{(+)}(3)$ and $SU^{(-)}(3)$ algebras. Again as in Sections 3 and 5 we have performed DSM calculaions using $H_Q^{(-)}$ by obtaining HF sp spectrum and then by angular momentum projection the ground $K=0$ band members for the six proton system. Now, using $T^{E2}=e_p\, Q^2_\mu(-)\;b^2$ where $b$ is the oscillator length parameter, 
with $e_p=1e$ and $b=2fm$, we obtain
\be
\barr{l}
Q(2^+_1)= -31\,e\,fm^2,\;\;\;\;\; Q(4^+_1)= -39\,e\,fm^2, \\
B(E2; 2^+_1 \rightarrow 0^+_1) = 232\,e^2\,fm^4, \;\;\;\;\;
B(E2; 4^+_1 \rightarrow 2^+_1) = 313\,e^2\,fm^4\;. 
\earr
\ee
However, DSM calculation with $H_Q^{(+)}$ and using the same $T^{E2}$ as above gives
\be
\barr{l}
Q(2^+_1)= 9\,e\,fm^2,\;\;\;\;\; Q(4^+_1)= 8\,e\,fm^2, \\
B(E2; 2^+_1 \rightarrow 0^+_1) = 19\,e^2\,fm^4, \;\;\;\;\;
B(E2; 4^+_1 \rightarrow 2^+_1) = 16\,e^2\,fm^4\;. 
\earr
\ee
Thus, as expected, $H_Q^{(-)}$ generates prolate shape and 
$H_Q^{(+)}$ generates oblate shape. On the other hand, if we change the $E2$ 
transition operator to $T^{E2}=q_2 Q^2_\mu(+)\,b^2$, then $H_Q^{(-)}$ gives $Q(2^+_1)= 
9\,e\,fm^2$ and $Q(4^+_1)= 8\,e\,fm^2$. Similarly, $H_Q^{(+)}$ gives $Q(2^+_1)= -31\,e\,fm^2$, $Q(4^+_1)= -39\,e\,fm^2$. As seen from these results, clearly when we use $H_Q^{(\baa)}$ 
with $(\baa)$ changing, the form of the $E2$ transition operator should not change. In SM literature \cite{SM-MM1,SM-MM2}, for $SU(3)$ limit the choice is correctly $T^{E2}=q_2 Q^2_\mu(-)$ so that $H_Q^{(-)}$ gives prolate and $H_Q^{(+)}$ oblate shapes.

\ed
\section*{References}
\begin{thebibliography}{}

\bibitem{El58-1} J.P. Elliott, Proc. Roy. Soc. (London) \textbf{A245}, 128 (1958).

\bibitem{El58-2} J.P. Elliott, Proc. Roy. Soc. (London) 
\textbf{A245}, 562 (1958).

\bibitem{Iac} F. Iachello and A. Arima, \textit{The Interacting Boson Model} (Cambridge University Press, Cambridge, 1987).

\bibitem{sm1} J. P. Draayer and K. J. Weeks, Ann. Phys. (N.Y.) \textbf{156}, 41 (1984). 

\bibitem{sm2}D. Bonatsos, I. E. Assimakis, N. Minkov, Andriana Martinou, S. Sarantopoulou, R. B. Cakirli, R. F. Casten and K. Blaum, Phys. Rev. C \textbf{95}, 064326 (2017).

\bibitem{sm3} C.L. Wu, D.H. Feng and M.W. Guidry, Adv. Nucl. Phys. \textbf{21}, 227 (1994).

\bibitem{sm4} D.J. Rowe, Prog. Part. Nucl. Phys. \textbf{37}, 265 (1996).

\bibitem{sm5} T. Dytrych, K. D. Launey, J. P. Draayer, P. Maris, J. P. Vary, E. Saule, U. Catalyurek, M. Sosonkina, D. Langr and M. A. Caprio, Phys. Rev. Lett. {\bf 111}, 252501 (2013).

\bibitem{sm6} K.D. Launey, T. Dytrych, G.H. Sargsyan, R.B. Baker and J.P. Draayer, Eur. Phys. J. Special Topics {\bf 229}, 2429 (2020). 

\bibitem{sm7} J. Cseh and K. Kato, Phys. Rev. C {\bf 87}, 067301 (2013).

\bibitem{sm8} J. Cseh, Eur. Phys. J. Special Topics {\bf 229}, 2543 (2020). 

\bibitem{Kota-book} V.K.B. Kota, \textit{$SU(3)$ Symmetry in Atomic Nuclei} (Springer Nature, Singapore, 2020).


\bibitem{Piet} P. Van Isacker, K. Heyde, J. Jolie and A. Sevrin, Ann. Phys. (N.Y.) {\bf 171}, 253 (1986). 

\bibitem{ib2} J.P. Elliott, Prog. Part. Nucl. Phys.  {\bf 25}, 325 (1990).

\bibitem{ib3} P. Halse, P. Van Isacker and B.R. Barret, Phys. Lett. {\bf B363}, 145 (1995).

\bibitem{ib4} H.Y. Ji, G.L. Long, E.G. Zhao and S.W. Xu, Nucl. Phys.  {\bf A658}, 197 (1999).
 
\bibitem{ib5} Y.D. Devi and V.K.B. Kota, Pramana-J. Phys. \textbf{39}, 413 (1992).

\bibitem{ib6} F. Iachello and P. Van Isacker, \textit{The Interacting Boson-Fermion Model} (Cambridge University Press, Cambridge, 1991).

\bibitem{ib7} R. Bijker and V.K.B. Kota, Ann. Phys. (N.Y.) \textbf{187}, 148 (1988).

\bibitem{ib8} V.K.B. Kota and U. Datta Pramanik, Euro. Phys. Jour. A \textbf{3}, 243 (1998).

\bibitem{KS-book} V.K.B. Kota and R. Sahu, \textit{Structure of Medium Mass Nuclei:
Deformed Shell Model and Spin-Isospin Interacting Boson Model} (CRC press, Taylor \& Francis group, Boca Raton, Florida, 2017).

\bibitem{Ko-17} V.K.B. Kota, Bulg. J. Phys. \textbf{44}, 454 (2017); arXiv:1707.03552 (2017).

\bibitem{KSP-19} V.K.B. Kota, R. Sahu and P.C. Srivastava, Bulg. J. Phys. \textbf{46}, 313 (2019).

\bibitem{JCP} J.C. Parikh, \textit{Group Symmetries in Nuclear Structure} (Plenum, New York, 1978).

\bibitem{EPJST} R. Sahu, V.K.B. Kota and P.C. Srivastava, Eur. Phys. J. Special Topics {\bf 229}, 2389 (2020).

\bibitem{Ko-irreps} V.K.B. Kota, arXiv:1812.01810 [nucl-th] (2018).

\bibitem{KM-1} S. Kuyucak and I. Morrison, Ann. Phys. (N.Y.) \textbf{181}, 79 (1998).

\bibitem{RMP} P. Cejnar, J. Jolie and R. F. Casten, Rev. Mod. Phys. {\bf 82}, 2155 (2010).

\bibitem{chaos1} D. Kusnezov, Phys. Rev. Lett. {\bf 79}, 537
(1997).

\bibitem{chaos2} A Shirokov, N A Smirnova and P Van Isacker, Phys. Lett. {\bf B434}, 237 (1998).

\bibitem{SM-MM1} G.F. Bertsch, The practitioner’s shell model (North-Holland, Amsterdam, 1972).

\bibitem{SM-MM2} P.J. Brussaard, P.W.M. Glaudemans, Shell model applications in nuclear spectroscopy (North-Holland, Amsterdam, 1977).

\bibitem{Draayer1} J.P. Draayer and G. Rosensteel, Nucl. Phys. {\bf A439}, 61 (1985).

\bibitem{Draayer2} G. Rosensteel, J.P. Draayer and K.J. Weeks, Nucl. Phys. {\bf A419}, 1 (1984).

\bibitem{Verg} J.D. Vergados, Nucl. Phys. {\bf A111}, 681 (1968).

\bibitem{AkiDra} J.P. Draayer and Y. Akiyama, J. Math. Phys. {\bf 14}, 1904 (1973).

\bibitem{Bon-1} G. Vanden Berghe, H.E. De Meyer and P> Van Isacker, Phys. Rev. C {\bf 32}, 1049 (1985).

\bibitem{Bon-2} N. Minkov, S.B. Drenska, P.P. Raychev, R.P. Roussev and D. Bonatsos, Phys. Rev. C {\bf 55}, 2345 (1997).

\bibitem{Bon-3} N. Minkov, S.B. Drenska, P.P. Raychev, R.P. Roussev and D. Bonatsos, Phys. Rev. C {\bf 60}, 034305 (1999).

\bibitem{Bon-4} P. Van Isacker, Phys. Rev. Lett. {\bf 83}, 4269 (1999).

\bibitem{tagg1} R. Sahu and S.P. Pandya, Nucl. Phys. A {\bf 414}, 240 (1984).

\bibitem{tagg2} R. Sahu, Nucl. Phys. A {\bf 501}, 311 (1989).

\bibitem{tagg3}	R. Sahu and S.P. Pandya, Nucl. Phys. A {\bf 529}, 20 (1991).

\bibitem{DrRo} J.P. Draayer and G. Rosensteel, Nucl. Phys. {\bf A386}, 189 (1982).

\bibitem{Scholten} O. Scholten, K. Heyde, P. Van Isacker, J. Jolie, J. Moreau and M. Waroquier, Nucl. Phys. {\bf A438}, 41 (1985). 

\bibitem{KM-3} S.C. Li and S. Kuyucak, Nucl. Phys. {\bf A604}, 305 (1996). 

\bibitem{Barr} A.F. Diallo, E.D. Davis and B.R. Barrett, Ann. Phys. (N.Y.) {\bf 222}, 159 (1993).

\bibitem{Warner} D.D. Warner, Phys. Rev. Lett. {\bf 47}, 1819 (1981).

\bibitem{Lipas} P.O. Lipas, E. Hammaren and P. Toivonen, Phys. Lett. {\bf B139}, 10 (1984). 

\bibitem{KM-2} S. Kuyucak and I. Morrison, Ann. Phys. (N.Y.) \textbf{195}, 126 (1989).

\bibitem{KY} Y.D. Devi and V.K.B. Kota, Nucl. Phys. {\bf A600}, 20 (1996).

\bibitem{Wu} H.C. Wu, A.E.L. Dieperink and O. Scholten, Phys. Lett. {\bf B187}, 205 (1987).

\bibitem{Hyde} K. Heyde, P. von Neumann-Cosel and A. Richter, Rev. Mod. Phys. {\bf 82}, 2365 (2010).

\bibitem{Richter} T. Beck et al, Phys. Rev. Lett. {\bf 118}, 212502 (2017).

\bibitem{JPD-1} O. Castanos, J.P. Draayer and Y. Leschber, Z. Phys. A - Atomic Nuclei {\bf 329}, 33 (1988).

\bibitem{DiBi} A.E.L. Dieperink and R. Bijker, Phys. Lett. {\bf B116}, 77 (1982).

\bibitem{Se-76} J. Henderson et al, Phys. Rev. C {\bf 99}, 054313 (2019).

\bibitem{JPD-2} D. Troltenier, C. Bahri and J.P. Draayer, Nucl. Phys. {\bf A589}, 75 (1995).

\bibitem{JPD-3} H.A. Naqvi and J.P. Draayer, Nucl. Phys. {\bf A516}, 351 (1990).

\bibitem{BPM} K.H. Bhatt, J.C. Parikh and J.B. McGrory, Nucl. Phys. {\bf A224}, 301 (1974).

\bibitem{Sun} Y. Sun, C.L. Wu, K.H. Bhatt, M. Guidry and D.H. Feng, Phys. Rev. Lett. {\bf 80}, 672 (1998).  

\bibitem{Zuk} A.P. Zuker, A. Poves, F. Nowacki and S.M. Lenzi,
Phys. Rev. C {\bf 92}, 024320 (2015).

\bibitem{Kr72}  H. Iwasaki et al., Phys. Rev. Lett. {\bf 112}, 142502 (2014).

\bibitem{Mag-1} A.K. Jain and D. Choudhury, Pramana-J. Phys. {\bf 75}, 51 (2010).

\bibitem{Mag-2} P.W. Zhao, S.Q. Zhang, J. Peng, H.Z. Liang, P. Ring and J.Meng, Nucl. Phys. {\bf A699}, 181 (2011).

\bibitem{Mag-3} C.M. Petrache et al., Phys. Rev. C {\bf 97}, 041304(R) (2018). 

\end{thebibliography}
